\documentclass[conference]{IEEEtran}
\IEEEoverridecommandlockouts
\usepackage{cite}
\usepackage{amsmath,amssymb,amsfonts}
\usepackage{algorithmic}
\usepackage{graphicx}
\usepackage{textcomp}
\usepackage{xcolor}
\usepackage{balance}       
\usepackage{graphics}      
\usepackage[T1]{fontenc}   
\usepackage{txfonts}
\usepackage{mathptmx}
\usepackage{color}
\usepackage{booktabs}
\usepackage{textcomp}
\usepackage{xcolor}
\usepackage{dirtytalk}
\usepackage{csquotes}
\usepackage{blindtext}
\usepackage{enumitem}
\usepackage{comment}
\usepackage{caption,subcaption}
\usepackage{multirow}
\usepackage{microtype}        
\usepackage{ccicons}          
\begin{document}

\title{Designing Indicators to Combat Fake Media
}

\author{\IEEEauthorblockN{Imani N. Sherman}
\IEEEauthorblockA{\textit{University of Florida} \\
shermani@ufl.edu
}
\and
\IEEEauthorblockN{Elissa M. Redmiles}
\IEEEauthorblockA{\textit{Microsoft Research} \\
eredmiles@gmail.com
}
\and

\IEEEauthorblockN{Jack W. Stokes}
\IEEEauthorblockA{\textit{Microsoft Research} \\
jstokes@microsoft.com
}
}

\maketitle

\begin{abstract}
The growth of misinformation technology necessitates the need to identify fake videos. One approach to preventing the consumption of these fake videos is provenance which allows the user to authenticate media content to its original source. This research designs and investigates the use of provenance indicators to help users identify fake videos. We first interview users regarding their experiences with different misinformation modes (text, image, video) to guide the design of indicators within users’ existing perspectives. Then, we conduct a participatory design study to develop and design fake video indicators. Finally, we evaluate participant-designed indicators via both expert evaluations and quantitative surveys with a large group of end-users. Our results provide concrete design guidelines for the emerging issue of fake videos. Our findings also raise concerns regarding users’ tendency to overgeneralize from misinformation warning messages, suggesting the need for further research on warning design in the ongoing fight against misinformation.

\end{abstract} 


\section{Introduction}
The rise of misinformation has been troubling academics and industry experts alike. Misinformation is associated with increasing political polarization and societal divisiveness and is considered by many as one of the most important problems currently threatening society~\cite{Chesney2019}. Even though propaganda and misinformation have existed long before the Internet, online technologies have accelerated and exacerbated the spread of such content~\cite{Anderson2017}. 

While the vast majority of prior academic research and industry interventions have focused on text-based misinformation, increasing concern is being raised about 
broader modes of fake media, including video, images and audio, 
that have either been subtly manipulated or generated entirely from scratch. Fake media broadly encompasses deep fakes, shallow fakes, and classic signal processing  methods. Deep fakes are generated using deep learning algorithms~\cite{Goodfellow14}, while the more prevalent shallow fakes rely on much simpler approaches such as photoshopping images or dropping video frames~\cite{Mervosh2019}. Signal processing methods can also be used to create new or alter existing media and can either used alone or in conjunction with deep fake algorithms~\cite{thies2016face} to create fake media.  
Such fake video media has been shown to be both extremely convincing and very difficult for people to detect in the wild, far more so than text-based media~\cite{dolhansky2020deepfake}. 

Four primary methods have been proposed to help identify and prevent users from falling for fake video and image content~\cite{roessler2019faceforensics++}:
\begin{itemize}
    \item \textit{Education}: Programs which focus on educating users about the problem of fake media and how to identify it~\cite{Barnwell2020}.
    \item \textit{Manual Investigation}: Careful training of media forensic analysts and journalists to identify and prevent the publication of fake media~\cite{Marconi18}. 
    \item \textit{Automated Detection}: Algorithmic-based (e.g., deep learning methods) detection of fake media~\cite{8630787,li2018exposing, 8630761, guera2018deepfake, da2012video, rossler2018faceforensics, zhang2018face, dolhansky2020deepfake}. 
\item \textit{Provenance-Based Authentication}: Systems that allow for cryptographically authenticating media content to its original source such as a newspaper or broadcaster~\cite{CAI,Amber,engl2020amp,ProofMode,Newman19}.
\end{itemize}

While exciting progress is being made toward automatically detecting and removing deep fake media, such approaches are far from ready to be released at Internet-scale~\cite{dolhansky2020deepfake}. Thus, the human must remain in the loop to interpret signals regarding the authenticity of video media content. A key question remains: how should we go about communicating to users the authenticity of the media to which they are exposed? While prior work in industry has explored how to communicate the authenticity of images~\cite{koren2019introducing}, for example by providing information about when and where the image was taken, no prior academic work to our knowledge has examined how to communicate the authenticity of \textit{video} media to end-users. 

In this work, we conduct a four-part, mixed-methods study as a first step toward understanding users' perceptions of fake media and developing user experience (UX) best practices to combat it. First, to gain context on people's existing mental models around different modes of fake media, we conduct a semi-structured interview study with 24 Americans from diverse sociodemographic backgrounds in which we probe participants concerns with media -- including image and video media as well as text media -- as well as their existing strategies for evaluating the truthfulness and authenticity of content. We find that users rely heavily on the source of video and image content in order to evaluate it, far more so than for text-based information. We thus focus on source-related misinformation indicators. Further, provenance systems -- which can inform the user whether a piece of content has been transmitted by or derived from the original source -- are the closest to being ready for widespread implementation and can provide concrete labels of media authenticity (vs. detection systems which provide a probability of authenticity, see Section~\ref{sec:rw:deepfakes} for more detail). Thus, we focus our remaining design steps on designing source authenticity (i.e., provenance) indicators for video media in order to align with users' existing mental models and best practice in industry. In the discussion, we review the ways in which our work can generalize to future work using automated detection mechanisms. 

\begin{figure*}[h]
    \centering
\begin{subfigure}{0.45\textwidth}
  \includegraphics[width=\linewidth]{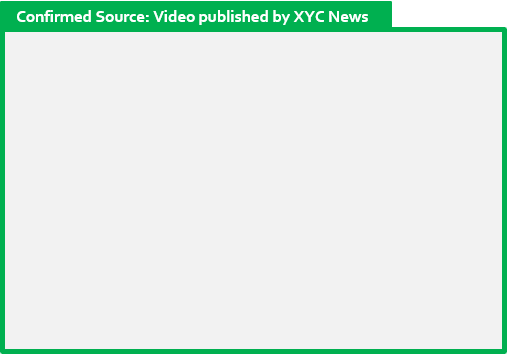}
  \caption{CLB:This warning uses the signal word "confirmed" and a green border.}
  \label{fig:1du}
\end{subfigure}\hfil
\begin{subfigure}{0.45\textwidth}
  \includegraphics[width=\linewidth]{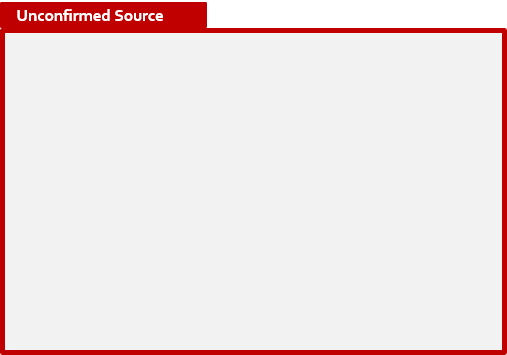}
  \caption{UcSB:This warning uses the signal word "unconfirmed" and a red border. }
  \label{fig:2du}
\end{subfigure}

\medskip
\begin{subfigure}{0.45\textwidth}
  \includegraphics[width=\linewidth]{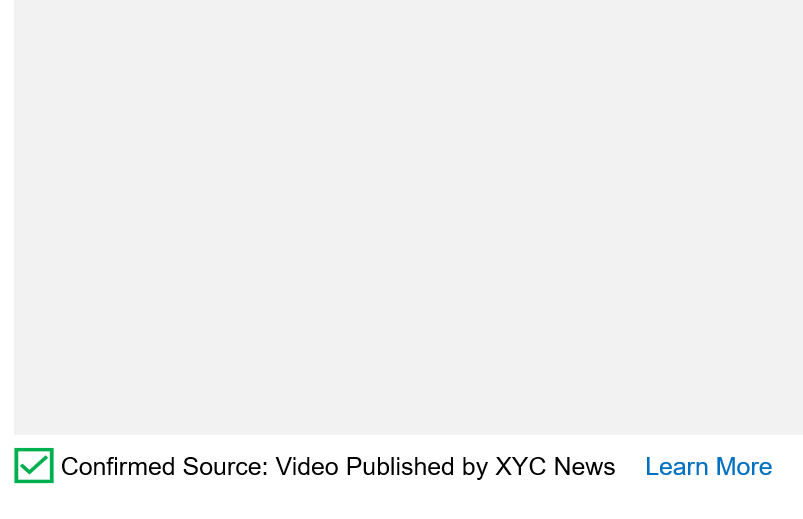} 
  \caption{CLC:This warning uses the signal word "confirmed" and a green check mark icon.  }
  \label{fig:3du}
\end{subfigure}\hfil 
\begin{subfigure}{0.45\textwidth}
  \includegraphics[width=\linewidth]{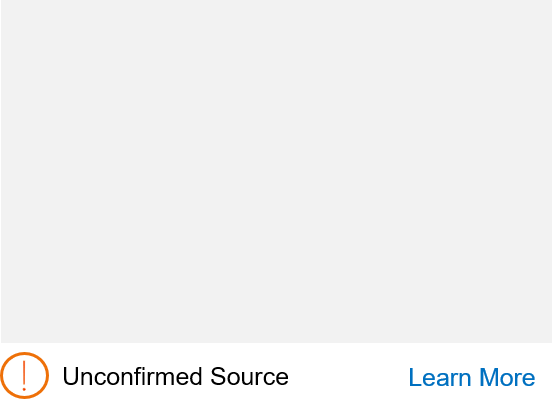} 
  \caption{UcSO: This warning uses the signal word "unconfirmed and an orange exclamation mark. }
  \label{fig:4du}
\end{subfigure}
\caption{The four most effective warning designs developed in this research. The full set of designs are shown in Appendix \ref{App:DesignImagesinSurvey} in Figures~\ref{fig:borderdesigns},~\ref{fig:greencheckmarks}, and~\ref{fig:exclamdesigns}.}
\label{fig:4of10designs}
\end{figure*}

Thus, second, situated in the context gained from our interview study, we conducted a series of participatory design studies with 19 Americans, again from diverse backgrounds, to develop a set of potential UX indicators for alerting users to the provenance of video content. Third, we refined these UX indicators through a series of expert reviews with five academic and journalistic experts from major news organizations and top academic institutions. Fourth, we quantitatively evaluate (n=1,456) a final set of proposed UX indicators (Figure \ref{fig:4of10designs} shows those that were found to be most effective), which were developed through participatory design and expert feedback.

In this work, we make the following contributions:

\begin{enumerate}
    \item \textbf{Differences in Misinformation Detection.} To our knowledge, we are the first to identify the differences in techniques that news consumers use to detect misinformation
    for different media types (text, image, video).
    \item \textbf{User-Centered Provenance Warning Design Guidelines.} Due to the nuances related to provenance warnings for videos, users and experts heavily emphasized the need for brevity, specificity, and simplicity through their feedback and suggested designs.
    \item \textbf{Over-generalization of Warnings.} Our quantitative survey results suggest that users often overgeneralize the meaning of misinformation warnings.
\end{enumerate}

Our results offer implications for future research on misinformation, suggest concrete designs and guidelines for warnings of video authenticity, and raise concerns around how users may overgeneralize in their interpretations of misinformation warnings.

\section{Related Work}
In this section we briefly review prior work on detecting deep fakes, on users' perceptions of media information, and on the design of warnings for misinformation and related technology applications.  

\subsection{Deep Fakes and Detection}
\label{sec:rw:deepfakes}
A face swap is one type of deep fake video where the face of someone in the original video is replaced with the face of a targeted person in the fake video using artificial intelligence~\cite{8630761,thies2016face}. Although this technology is often used by social media users to create unique images and videos, this technology has also been used to create fake sexually explicit videos and has the potential to be used for other malicious activities. Due to this, researchers have begun to create deep fake video detection software that determines authenticity based on blinking~\cite{8630787}, face warping artifacts~\cite{li2018exposing}, and other techniques~\cite{8630761, guera2018deepfake, da2012video, rossler2018faceforensics, zhang2018face}. However, the state-of-the-art fake video detection systems are at best able to detect when content is authentic -- for example through provenance-based approaches~\cite{CAI,Amber,engl2020amp,ProofMode,Newman19} -- rather than when content is manipulated. Thus, we focus our design studies on developing indicators of authenticity rather than manipulation, as described in Section~\ref{sec:designstudy}. 

\subsection{User Perceptions of Media} 
Prior work has found that people may use multiple heuristics to identify misinformation. Multiple studies have found evidence that the source of the information is one such heuristic. This heuristic includes the reader's perception of the trustworthiness of the source itself~\cite{wagner2019reception, flintham2018falling}, the appearance of the website on which the information is shown~\cite{fogg2003users}, and whether the information confirms or aligns with what the reader already believes~\cite{metzger2010social,guess2019less}. In addition to looking at the source's appearance, the bias in reporting, and alignment with current beliefs, readers also look to see if other sources report the same information or if the information is believed by others~\cite{zaryan2017truth}.

On social media, readers also have additional context that prior work finds they leverage to inform their media perceptions. Specifically, the Media Insight Project found that the credibility of information posted on social media is influenced by who posts the information~\cite{sterret2018shared}. This is corroborated by other studies that found that Twitter users rely on the profile picture, name, bio, and the expertise of the person that posted the content to determine the validity of content~\cite{lee2013tweet, morris2012tweeting,geeng2020fake}. Also, similar to previous work on online news in general~\cite{metzger2010social}, Alcott et al.~\cite{allcott2017social} found that people were more likely to believe ``fake news'' stories on social media that favored the candidate they liked during the 2016 election.

Prior work by Sundar et al. proposes a theoretical framework for understanding end-user perceptions of fake news consisting of four cues - Modality, Agency, Interactivity, and Navigability - that users rely on to determine the credibility of the information~\cite{sundar2008main}. The first is Modality cues, where credibility is determined based on the structure of the medium providing the content. The second is Agency cues where credibility is based on the source of the information. The third is interactivity cues, where the actions one can take with the information is linked to credibility. The last is navigability where the user's ability to navigate the website of information impacts credibility. 

Finally, a limited body of prior work has explored these questions with participants in other countries, largely finding similar results. Similar to the findings of Wagner et al. in the U.S.~\cite{wagner2019reception}, Wasserman and Madrid-Morales~\cite{wasserman2018new, wasserman2019exploratory}, who surveyed adults in Kenya and Nigeria, found that participants had a low level of trust in the media and believed they received a high amount of misinformation. Valenzuela et al.~\cite{valenzuela2019paradox} surveyed residents in Chile and found that those who were politically engaged were more likely to share misinformation. Duffy et al.~\cite{duffy2019too} interviewed Singaporeans and found that people did not always know when they received fake news, but were open to sharing information if it appeared to be something that would benefit others. On the other hand, the emergence of fake news has made them more cautious about sharing and believing everything they read. Tandoc~\cite{tandoc2019tell} interviewed college students in Singapore who rated their Facebook friends as more credible than a news source. However, when motivation was high, the opposite behavior was exhibited and students rated the news source as more credible than their Facebook friends.

Thus far, research has shown that people's perceptions of the trustworthiness of media are influenced by their existing beliefs and the source of the media -- where source encompasses both elements of the news publisher, the UX of the news itself, and who shared the news. However, all of this prior work of end-user perceptions of media has either not explicitly addressed media mode (e.g., text vs. image vs. video) or have focused specifically on text media~\cite{geeng2020fake}. Given that participants may reasonably engage with different types of misinformation in different ways, and due to the growth of video misinformation especially, our interview study explicitly focuses on user experiences with different modes of misinformation. Additionally, prior work has focused exclusively on the truthfulness or trustworthiness of the content rather than provenance, e.g., whether the content itself was manipulated, despite the fact that provenance is among the most technically feasible methods for detecting fake news. Thus, in this work we explicitly explore provenance. 

\subsection{UX Indicators for Misinformation}
\label{sec:rw:misinfoux}
Prior work in HCI and the social sciences has focused on methods for: actively dissuading end-users from engaging with misinformation typically using warnings~\cite{dias2020emphasizing,ozturk2015combating,lazer2018science,ecker2010explicit,clayton2019real,lewandowsky2012misinformation,garrett2013promise,yaqub2020effects} but occasionally also via games and other educational methods~\cite{grace2019factitious}; identifying which content should be fact-checked~\cite{babaei2019analyzing,liu2018early,lazer2018science,shu2017fake,zhang2018structured}; and correcting inaccurate beliefs~\cite{bode2018see,porter2018sex}. We focus on the first body of related work as it is most relevant to the work presented here. 

The majority of prior work suggests that fact-check warnings can lead to small but significant improvements in the accuracy of users beliefs~\cite{ozturk2015combating,lazer2018science,ecker2010explicit,clayton2019real,yaqub2020effects,seo2019trust,ross2018fake}. The literature finds that misinformation warnings are most effective when they are immediate~\cite{lewandowsky2012misinformation,ozturk2015combating} (e.g., at the time of first exposure to the information)\footnote{Earlier prior work argues that immediate correction may, however, increase the risk of backfire effects~\cite{garrett2013promise}.}, presented in context (e.g., in close visual proximity to the content)~\cite{ozturk2015combating}, and are specific~\cite{ecker2010explicit,clayton2019real,ross2018fake} (e.g., indicate to the user exactly what is problematic about the content). 

Fact-checking is an art of persuasion: fact-checks are typically conducted on politicized content and thus misinformation warnings or direct efforts to correct incorrect beliefs risk back-fire effects in which the attempt to correct may lead to users' strengthening their belief in misinformation~\cite{garrett2013promise,guess2018does,wood2019elusive,nyhan2019taking}. 

While the vast majority of prior work on user perceptions of misinformation and credibility cues have focused on text content, a small body of prior work has focused on image media. The New York Times' News Provenance Project investigated ways to provide the context of photos and ensure the context remains with a photo as it is shared across the internet~\cite{koren2019introducing}. After interviewing 34 individuals with various political leanings about their media needs, their results suggested that users want the context in a simple way so they can form their own opinion. The researchers propose using blockchain technology to provide users with the image source, location, date published, and a short summary on what the photo captures.

No prior work, to our knowledge, has explicitly considered the design of indicators for provenance and especially provenance of video content. Our work takes a first step toward filling this gap.

\subsection{UX for Warnings}
The user experience literature also discusses the use of indicators to guide user behavior outside of the misinformation domain. For example, researchers in the usable security community have reviewed the use of warnings for browsers~\cite{felt2016rethinking, clark2013sok, thompson2019web, roberts2019you, wu2018your, egelman2008you, schechter2007emperor, stojmenovic2018they, yu2016exploring, reeder2018experience,redmiles2017you}, apps~\cite{abu2017obstacles, stavova2018experimental,shermanyou}, and physical spaces~\cite{scaife2019kiss, portnoff2015somebody}. Research-based guidelines for warning designs encourage removing visual clutter to improve warning detection, including consequences for non-compliance, use of a signal word, and clearly stating the hazard~\cite{wogalter2002based}. Investigations into the effectiveness of SSL warnings, which alert users when websites may not be as they appear or when their connection is unsecured, also found that hiding technical details~\cite{nodder2005users} and providing clear default choices~\cite{egelman2008you} reduce the chances that readers simply click through the warning without reading it~\cite{felt2014experimenting, akhawe2013alice}.
\section{Interview Study: User Experiences with Different Modes of Misinformation}
\label{sec:interview}
In June 2020, we conducted 24 semi-structured interviews with Americans from diverse sociodemographic backgrounds to understand their experiences with text, image, and video misinformation. These interviews provide context in which we situate our design studies (Section~\ref{sec:designstudy}). As aforementioned, little prior work has investigated experiences with image and video misinformation and thus it was necessary for us to conduct our own contextual investigation.

In this section, we review our methods and the results of our interviews; limitations of all studies are discussed in Section~\ref{sec:limitations}. All methods were approved by our institution's ethics review board.

\subsection{Methods}
\label{sec:interview:methods}
Our interviews were conducted via telephone or video conference due to COVID-19 restrictions. Interviews lasted for approximately 60 minutes, and all participants were compensated with \$20 Amazon gift cards for their participation.

\subsubsection{Interview Protocol}
Our interview protocol was organized into four sections. 
The first three sections covered participants' experiences with text, image, and video media, respectively. Each of these sections probed whether participants have ever had concerns about the truthfulness, accuracy or credibility of the text, image, or video information they receive. The interview probed what particular information they were most concerned about and what their concerns were, as well as how they went about assuaging those concerns (e.g., how they evaluated information) including asking the participant to walk through a concrete example of how they evaluated information, if they did so. The interview evaluated whether the participant had particular sources that they always or never trusted, their perceptions of information they saw on social media in particular, as well as whether they had concerns about non-news information (e.g., health or entertainment information). The interview also probed whether the participant had ever had any negative information experiences (e.g., falling for a piece of misinformation or manipulated media). 

Finally, the fourth section of the interview asked whether there was anything that the participants' trusted sources could do that would lead to the participant no longer trusting them and  whether the participants' approach towards text, video and/or image news had changed in the past 10 years.

We refined our interview protocol for clarity through a pilot session attended by two researchers in addition to the interviewer before proceeding with our interviews.

\subsubsection{Recruitment}
We aimed to recruit a demographically diverse set of participants whose demographics roughly matched those of the U.S. population. To do so we used two methods: we quota sampled 300 responses to a recruitment screening survey that collected respondents' demographics through the survey firm Cint, ensuring that we received a survey sample that was representative of U.S. demographics on age, gender, education, income, race, and geographic region. Respondents were compensated according to their agreement with Cint. We then contacted survey respondents who indicated interest in our full interview on this survey, attempting to balance participant demographics as we scheduled and interviewed respondents. 

Through this process, we found that younger respondents, those under 30 years of age, were less likely to respond to our requests to schedule an interview, despite initially indicating interest. Thus, to ensure a balanced interview population we posted an advertisement on Craigslist in our city of residence\footnote{Craigslist only allows one advertisement for one study to be posted, and thus we could not recruit in multiple cities.} using the same screening questionnaire to recruit interview participants and balance diversify our participant pool. See Table \ref{table:DemoInterview} in Appendix \ref{App:InterDemo} for the final demographics of our participants. 

\subsubsection{Analysis}
We analyzed our interview findings through open-coding \cite{burnard1991method}. Three researchers examined a randomly selected 20\% (5) of the interview transcripts to independently develop a codebook. The researchers then met to reach agreement on a final codebook. Two of the researchers then coded a newly selected 20\% of the interview transcripts and achieved $\kappa$=.67. Given this ``substantial'' agreement~\cite{landis1977application}, and in line with prior work \cite{lombard2002content}, one researcher proceeded to code the remainder of the transcripts (including the original five used for codebook creation).  

Consistent with our qualitative methods and best practice guidelines for qualitative research, we report our results purely qualitatively -- describing whether most, many, some, or a few -- participants reported a particular code~\cite{mcdonald2019reliability, sandelowski2001real}.

\subsubsection{Quantitative followup}
\label{sec:interview:meth:quant}
In order to supplement this qualitative lens and more broadly evaluate awareness of provenance issues (e.g., deep fakes, manipulated images), we conducted a brief follow up survey on Amazon Mechanical Turk (n=3,552)\footnote{Survey and all procedures discussed in this paper were IRB approved. Crowd workers were paid \$0.30 for the 2 minute survey task to ensure at least \$8/hr hourly wage. The participant demographics can be found in Table \ref{table:DemoDeepSurvey} in Appendix \ref{App:InterDemo}.} to ask a larger number of participants about: whether they had ever heard of a ``deep fake'' video (and if so, how they would define that term and whether they had seen one) and whether they had ever heard of (and if so, ever used) reverse image search. We also included a final question ``Your answer to this question will not affect your compensation. Did you do an online search (such as using Google) to help answer the questions we just asked you?'' 
See Section~\ref{sec:limitations} for a discussion of the use of Amazon Mechanical Turk (MTurk) workers. We analyzed our data through descriptive statistics and report our results below.

\subsection{Results}
In this section we report on the results of our interview study and follow up survey. These results provide (a) a context in which to situate the design of our indicators, and (b) the first, to our knowledge, comparative investigation of users' perceptions and experiences with different \textit{modes} of misinformation: video vs. image vs. text.

\subsubsection{The role of provenance in various media types}
Most participants mentioned various fact-checking methodologies reported in prior work \cite{zaryan2017truth,fogg2003users} such as triangulation (i.e., checking whether multiple sources showed the same information), using their perception of the information source, and using their intuition regarding the validity of the content itself. However, our work is the first, to our knowledge, to identify that participants rely on a different set of methods for each media type. To evaluate text media, most participants used triangulation, source and intuition; while context, intuition and triangulation were used by those who had a method for image evaluation. While few participants had an established method for evaluating video media, those who did relied on triangulation and source. For example, the following statements are from two participants when asked how they evaluate text and video, respectively:

\begin{displayquote}
\say{The first thing I do is look at the source, the URL and who the author is, that kind of thing. And if that checks out then I'll probably look on something like Reddit which is like
a crowd source tool, which you can have people chime in and say, they're listening about it,
and whether they think it's true or not. So I think the combination of looking at the source
of the article. I think I can get a better picture of whether it's correct or not.}  -P15 (Male, 20s)
\end{displayquote}

\begin{displayquote}
\say{Obviously I'll supplement [the video] with the actual articles or see if anyone else posted other videos. I can go around the Internet and see, "Oh, does anyone have a different angle of this, or does anyone have any other... ? Maybe three people video recording it," or something along those lines. I definitely try to go back and see, is there something I can go do to get a better vantage point or different opinion or something like that.} - P16 (Female, 40s)
\end{displayquote}

Although all participants had a predefined process for evaluating text content, some did not have established heuristics for evaluating of images, and many did not have heuristics for evaluating video because they felt these forms of media were too difficult to evaluate. For example, one participant responded quickly with the following when asked how they fact-check text information.

\begin{displayquote}
\say{Google. I Google everything.} -P9 (Male, 60s)
\end{displayquote}

Then, this same participant responded with the following when asked about evaluating images.

\begin{displayquote}
\say{Yes. You can really tell if the photo's photoshopped if you really know there's no way in hell that's real. There's no way in hell that's real, that's photoshopped. I get a chuckle out of that.} -P9 (Male, 60s)
\end{displayquote}

However, they did not have a method for evaluating video.

\begin{displayquote}
\say{No. I'm not able to. The only way I could check would be to take somebody else's word. You understand what I'm saying? I can't check. If you're going to check something, you've got to physically check it. You. If you're checking on the accuracy of something and you're taking somebody else's word for it …That's somebody else's word.}-P9 (Male, 60s)
\end{displayquote}

Our quantitative results suggest that some participants' struggle to identify heuristics for checking video or image content may be related to a lack of awareness of manipulated video media and standard methods for identifying manipulated image content such as reverse image search. We surveyed 3,552 crowd workers and found that 51\% of respondents had heard of deep fake videos and only 39\% had heard of reverse image search. When asked if they had searched for answers when completing our questions about these terms 34\% of respondents who said they heard of deep fake videos admitted to looking up the term and 26\% admitted to doing so regarding reverse image search. Accounting for those who looked up the terms to answer our question, 33\% of participants had heard of deep fake videos while only 30\% had heard of reverse image search.

We also asked participants which media type they were most and least concerned about containing misinformation. Although there was no clear consensus for which type was most or least concerning for participants, the explanations were similar. Most participants were concerned about videos and images being edited or taken out of context and the difficulty associated with evaluating them. Additionally, participants viewed text as necessary for understanding the context and easy to fact-check. For example, participants stated the following:

\begin{displayquote}
 \say{Probably [more concerned about misinformation in]the actual text because, like I said, you can't believe everything you see, right?} -P19 (Male, 40s) \\
 
 \say{I feel like a lot of people don't necessarily want to read, but if they're scrolling through social media or something, they might be more likely to watch a video...I mean, you might still have false information [in text], but since videos and photos can be altered so much, I would say I'm less concerned about the text of a news article.} -P17 (Female, 30s) \\

\say{Images are the ones I'm most concerned about, but they just tend to be more accurate than texts articles. I don't know. I think that text articles are easier to check, so even if they are inaccurate but I can find out, whereas images I don't know how often they are accurate or not, just because it's harder to check.}-P15 (Male, 20s)
\end{displayquote}

\subsubsection{Use of Trusted Sources}
Most participants referenced the credibility of a source to help determine if the information being presented was trustworthy. Most participants had a list of trusted sources that they often referenced. For example, when we asked a participant about their trusted news sources they stated the following:

\begin{displayquote}
\say{Yeah, Seattle Times is a good one. New York Times. What else? Washington Post is a good one. MSNBC, of course, CNN. Those are all ones that I really trust most of the time. I don't think I've ever had a fake story posted on there.} -P19 (Male, 40s)
\end{displayquote}

However, most participants were unable to identify news sources they \textit{did not} trust, they were receptive to information from sources outside their trusted list, as long as they perceived the information as credible. For example, one participant stated the following:

\begin{displayquote}
\say{I don't think it's to a point of never [trusting a particular source]. I just read all types of different sources, and if I find something that is a little funny or interesting, or not believable, I just kind of look at other places see if they're the same, or if something is small, I just let it go. But it's not to a point of never [trusting somewhere].} -P13 (Female, 20s)
\end{displayquote}

\subsubsection{Social Media Posts}
While participants mentioned getting a high volume of visual news information through social media, they were very skeptical about the information they received. Even though participants were willing to read and receive information from unfamiliar news sources, this openness did not extend to social media posts, unless those posts were sharing information from a source they already explicitly trusted. For example, three of our participants stated the following:

\begin{displayquote}
\say{I'm not very trusting [of] anything that's shared over social media. If a video is from a local news outlet that I trust, then I'm more keen or more trusting on that.} -P23 (Male, 30s)
\end{displayquote}

\begin{displayquote}
\say{[On] social media, [I'm] probably the most [concerned]...Because you don't know where people are pulling that stuff from.} -P11 (Female, 40s)
\end{displayquote}

\begin{displayquote}
\say{Yeah, because I mean, I appreciate [social media] for photographs coming from people who are at the scene without being employed by the [news or] anything like that, but then again you get the same question of whether the image is correct or not. Whereas you don't have that doubt with the news agency.}  -P15 (Male, 20s)
\end{displayquote}

\subsubsection{Misinformation Impact}
Having experienced misinformation in various forms in the past, news consumers have adopted questioning the authenticity of text content, as found also in prior work \cite{amazeen2019reinforcing, edgerly2020audiences}.

\begin{displayquote}
\say{I used to believe everything. Now it's the complete opposite. First, I don't believe it unless I verify...there's so much stuff that's circling around that you don't even know at this point. It's like you have to make your own judgment now. You can't trust anything you're seeing.} -P8 (Male, 30s)
\end{displayquote}

\begin{displayquote}
\say{Honestly, I [started checking after] the last election and all the information that came out about whether or not Russia had something to do with Trump winning the election...seeing news about that is what has made me start to really question things.} -P17 (Female, 30s)
\end{displayquote}

Our results suggest that many are applying this same scepticism to image and video media, but may not know how to investigate this content beyond looking for obvious cues. For example, some of our participants stated the following:

\begin{displayquote}
\say{I've always had videos in my life, videos that are online or on TV and things like that. I don't know if that's the case from before, but I mean, it's always been more of the sources that I would trust on TV, and less of a source I would trust anywhere else, just because it's easy to make a video and movie editing has always been a thing.} -P15 (Male, 20s)
\end{displayquote}

\begin{displayquote}
\say{ I think I'm way more skeptical, just because technology has changed so much. 10 or 20 years ago, I would look at a picture and that's what I'm seeing, where now, I have to kind of stop and think, is this really what I'm seeing?}-P10 (Female, 40s)
\end{displayquote}

\begin{displayquote}
\say{Yeah, I think it's pretty much the same as with articles or images or anything like that. I want to consider the source that the person is sharing from, before I know whether it's... This is going to sound rude, but even worth my time.}-P13 (Female, 20s)
\end{displayquote}

\section{Design Study: User-Driven Development of\\Video Authenticity Indicators}
\label{sec:designstudy}

In July 2020, we conducted 19 participatory design sessions with Americans from diverse sociodemographic backgrounds to develop a set of potential UX indicators for communicating media provenance for video content (e.g., whether the video being shown to the user was actually from the claimed source). We focus specifically on provenance for two reasons. First, as described in more detail in Section~\ref{sec:rw:deepfakes}, the most promising technological innovations on combating deep fakes focus on provenance. These are the technologies beginning to be deployed by major news organizations~\cite{originIbc,BBCTrust32:online}, and thus focusing on provenance has the most immediate impact. Second, as evidenced by our interview study results, end-users already rely heavily on the source of content, especially for image and video media. Thus, conducting research on provenance is in alignment with users' pre-existing mental models and practices.

Below, we describe our methods and the results of our design interviews. We also describe our follow up expert design reviews of the participatory design-produced indicators, and the results of those reviews. All methods were approved by our institution's ethics review board.

\subsection{Methods}
Our participatory design sessions were conducted via video conference due to COVID-19 restrictions. Design sessions lasted for approximately 60 minutes, and all participants were compensated with \$20 Amazon gift cards for their participation.

\subsubsection{Participatory Design Protocol}
Design sessions consisted of three parts: in the first, participants sketched notification designs while in the second part they gave feedback on other designs (see Figure \ref{fig:dpborder}, \ref{fig:dpNPP} and \ref{fig:dptwitter} in Appendix \ref{App:designprobes}). In the third, participants instructed the researcher on how to create their ideal design using PowerPoint. A subset of participant designs are shown in Figure \ref{fig:partdesign2} of Appendix \ref{App:designprobes}.

We started the design sessions by asking participants to ``draw some messages that you think would be able to inform someone looking at a video about whether it really came from the original source / the source stated on the video''. We conducted these drawing exercises in the context of a scenario, to improve concreteness and maximize the validity of the sketches: ``Imagine you are reading an article on the Blue Bear Times website, and you see a video associated with the article that appears to have been created by Blue Bear Times. Please draw what you would expect to see, if our technology [was unable to verify/was able to verify] that the Blue Bear Times was the source of the video/ created that video.'' We had a similar set of scenarios to prompt participant sketching in the context of a video appearing in social media. 

\begin{figure}[t]
\centering
  \includegraphics[width=.8\linewidth]{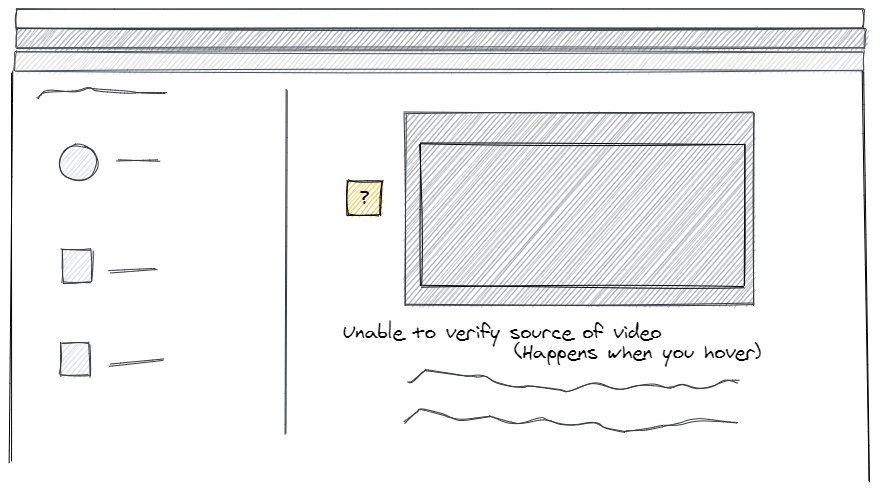}
\caption{This figure shows a screenshot of participant P7's unverified source warning design made using the Excalidraw template we provided. This warning includes a yellow box with a question mark to the left of the video. When the user hovers over the box, a message appears that says ``Unable to verify source of video."}
\label{fig:ExcalExample}
\end{figure}

Participants sketched either within a drawing template that we provided (Figure \ref{fig:ExcalExample}) using Excalidraw \cite{exdr} or told us what to sketch and watched and revised as we sketched it.\footnote{Due to COVID19 restrictions we were unable to conduct in-person design studies.} 
After completing their sketches, participants evaluated a set of potential provenance notification designs that we created based on existing fake news notifications (see Appendix~\ref{App:designprobes} for the designs). To avoid desirability bias (participants softening their feedback because they knew we created the designs), we told participants that we were not apart of the design process but were hired to get user feedback. For each design, we asked: ``What do you immediately notice about this warning/label?'', ``What do you like about it?'', ``What do you dislike?'', and ``How can it be improved?''
Finally, after participants viewed all  designs, we reviewed which designs, design elements, and phrasings were their favorites among their own designs and the other designs they saw.

\subsubsection{Recruitment}
We used the same recruitment approach as described in Section~\ref{sec:interview:methods}. See Table \ref{table:DemoPartDesign} for the final demographics of our design study participants. 

\subsubsection{Analysis}
We chose to do participatory design studies to maximize inclusion of diverse perspectives in the design of our indicators~\cite{mcdonald2019reliability,marwick2017nobody}. 
In order to avoid oppressing these perspectives~\cite{bellini2018feminist} or reducing our sensitivity to nuances in the participatory design data by too restrictive of a formal coding process, we followed the best practice guidelines of McDonald et al.~\cite{mcdonald2019reliability} . To this end, we used a field notes approach to analyze our results and reached consensus on our findings through researcher discussion of the results rather than open-coding with intercoder agreement.

\subsubsection{Expert Design Reviews}
We also conducted five expert design reviews with experts who had major mainstream newspaper and/or academic journalism backgrounds. We met with each expert for 30 minutes and solicited their feedback on the set of designs most preferred by our participants. Each expert was compensated with a \$50 Amazon gift card.

All reviews were attended by three members of the research team, and the research team discussed the results of the session, reaching consensus on the feedback. In the results section below, we qualitatively describe the feedback provided by these experts. 

\subsection{Results}

\subsubsection{Participant Designs.} Overall, participant's designs consisted of the colors red or yellow for notices about un-authenticated media and green for notices indicating authenticated media (see Appendix \ref{App:designprobes}). Participants used either the phrase \textit{(un)verified} or \textit{(un)known} source to indicate the authentication status of the media. Participants used either a check mark or green block of color (for authenticated media) and an exclamation point  or red block of color (for unauthenticated media). A few participants used interstitial warnings or messages imposed over the video. 
However, the majority of the participants' designs included a check mark or an exclamation point icon, as appropriate, and a phrase below the video that included the word \textit{verified} when the video was authenticated  and \textit{unverified} or \textit{not verified} when the video could not be authenticated. 
(see Figure \ref{fig:partdesign} and Appendix~\ref{App:designprobes}).

\subsubsection{Design Feedback} Participants' discussions of their own designs, others designs, and the design probes, as well as experts' feedback on participants' designs can be summarized into four suggestions. We discuss each suggestion below and how it impacted the designs we chose for quantitative testing.

\textbf{Include a simple and clear message.}
As shown in Figure~\ref{fig:dpNPP} in Appendix~\ref{App:designprobes}, some of the design probes included detailed information, such as date, location and subject, as suggested by prior work on image media~\cite{koren2019introducing}. Participants mentioned that while they understood why this information could be useful, they just wanted to read a simple phrase that was easy to understand and would not cause them to have to pause for more than a few seconds to read and react to the message, so as not to interrupt their flow of consuming content. Thus, they preferred that additional contextual information be placed deeper in the UX, such as behind a \textit{Learn More} link. 
For example, one participant stated:

\begin{displayquote}
\say{I just think that [all the detailed information] would be a bit too much.... I like the learn more [instead of the additional information]... I'd rather just [know that] the video [is] verified or unverified.} -P2 (Female, 30s) 
\end{displayquote}

\begin{displayquote}
\say{I feel like this [context about the video] almost takes away from the content itself... That's just a lot of information. I would prefer something that I could just quickly glance at to see if something is verified or not.} -P15 (Female, 20s)
\end{displayquote}

After receiving this feedback consistently from the participants, we decided to focus on testing warnings with short phrases that included the \textit{Learn More} option. It is however important to note that participants may be unlikely to access the additional information, as shown by recent work examining participants' real practices when exploring news content on social media~\cite{geeng2020fake}.

Additionally, while most participants used the word \textit{verified} in their warning designs before and after seeing the design probes, 
 perhaps because this is standard language used in social media for authenticated users, experts were concerned about the use of this word. When evaluating a participant design that contained the word \textit{unverified}, Expert A stated:

\begin{displayquote}
"I don't know that this is the best word choice. It makes me want to ask... Well who says its unverified?... unconfirmed is a little more exact"-Expert A
\end{displayquote}

We received similar feedback from the other expert reviewers in favor of the word \textit{confirmed}, which prompted us to quantitatively test the word \textit{confirmed} against the signal word \textit{verified}, which experts worried would be over interpreted: 
\begin{displayquote}
\say{My main concern is that people will read too much into it and might take it to mean too much more than what it actually means.}-Expert C
\end{displayquote}

\textbf{Tell me why it is unauthenticated.}
Similarly, participants wanted to be able to find information specifically about why a video might be unverified or unconfirmed. While they did not want this information as part of the main warning, they wanted this information to be available if desired and suggested it be added to some of the designs that did not have it (see Figure~\ref{fig:dp10} in Appendix~\ref{App:designprobes}).

\begin{displayquote}
\say{Not being able to determine the source is different from we don't think this is a good source... I want to be clear about the intention...Something that is not a verifiable source is based on we don't know the source... Is there a learn more?... Ok, so if someone comes across this, and they'd want to know how come... how do you help answer that?} -P13 (Male, Over 60)
\end{displayquote}

The expert reviewers agreed that having a \textit{Learn More} button would be beneficial. They believed that participants might have questions about the authentication process and adding the \textit{Learn More} button would allow viewers to understand the warning better.

\textbf{Be specific.}
In addition to keeping warnings short and clear, but with additional context available through a \textit{Learn More} button, participants and experts wanted to ensure that warnings were as specific as possible; supporting prior work showing that specific warnings are more effective~\cite{ecker2010explicit,clayton2019real,ross2018fake}. 

\begin{figure} [ht!]
    \centering
   \begin{subfigure}{0.45\textwidth}
 \centering
  \includegraphics[scale=.45]{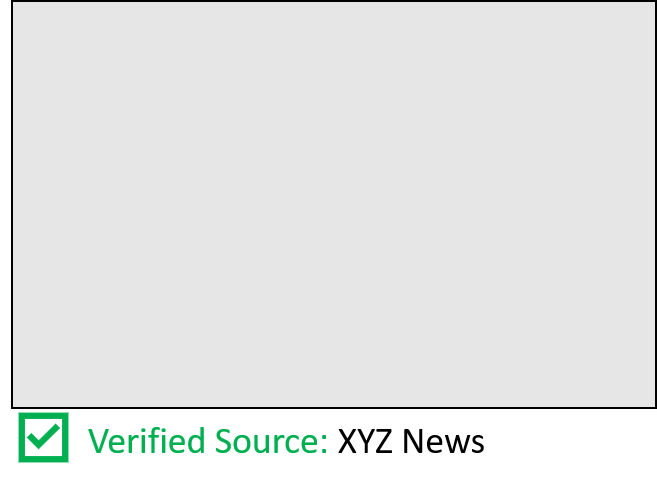}
\end{subfigure}\hfil
\begin{subfigure}{.45\textwidth}
   \centering
  \includegraphics[scale =.45]{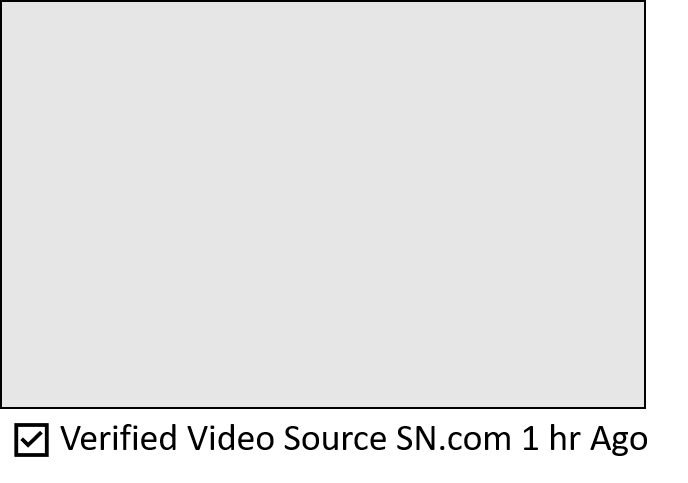}
\end{subfigure}
    \caption{These are screenshots of two warning designs created by participants during the study. Additional participant created designs can be found in Appendix \ref{App:designprobes}.}
    \label{fig:partdesign}
\end{figure}

A few of the designs we showed participants announced when a video source was confirmed but did not clearly state the source itself (see Figure \ref{fig:dp14} in Appendix \ref{App:designprobes}). Participants' responses to these design probes and some of their own designs (see Figure \ref{fig:partdesign}) suggest providing explicit information regarding the source of the video to make it clear to the reader what had been confirmed or verified. 

\begin{displayquote}
\say{So it's saying the source of this video is known. Ok. Why are you telling me the source of this video is known? Like ,who is it? Like, why wouldn't you just say it? Like, put that in there with it.} -P19 (Female, 30s)
\end{displayquote}

Our expert reviewers agreed with this sentiment and encouraged making the warnings highly specific. Instead of simply saying ``Verified: XYC News'', Expert B suggested using ``Verified Source: This video was published by XYC''. We implemented a shorter version of this statement using ``Verified Source: Video published by XYC News'' (see Figure~\ref{fig:1du}) and compared this against the even shorter ``Verified: XYC News'' in our quantitative tests presented in the next section.

\textbf{Do not interrupt typical video viewing flow.}
Finally, participants felt strongly that the warnings should not interrupt their video viewing flow. This is in line with prior work finding that people want to be empowered to evaluate content, but given the autonomy to make their own choices about what to consume~\cite{koren2019introducing}. 
Some design probes showed passive warnings (e.g., a message below the video) while others showed interstitial warnings that required action before the viewer could watch the video. Although our participants understood why these design characteristics might be necessary, they preferred a warning that easily captures the viewer's attention but is easy to ignore when watching a video. Our participants did not want to be distracted by the warning when trying to watch their video of choice. In response to Figure~\ref{fig:dp4}  (Appendix~\ref{App:designprobes}), a participant stated the following:

\begin{displayquote}
\say{Most of these [warnings with colored borders] are drawing way too much attention to themselves. This brings me back to the ideas that I drew which were unobtrusive [see Figure~\ref{fig:ExcalExample}]} -P7 (Male, 20s)
\end{displayquote}

Relatedly, participants wanted to avoid requiring additional clicks to watch a video after the warning is displayed. Our participants stated that the goal of the warnings was to notify the user of the authenticity of the media (that it was unmodified from the news source) and not to notify the user of anything about the information itself. Since the information presented could still be accurate, even though the source was not authenticated, there is no reason to prevent users form watching the video. For example, when we showed the design displayed in Figure~\ref{fig:dp12} (Appendix~\ref{App:designprobes}), P5 immediately responded by saying:

\begin{displayquote}
\say{Why are you making the customer click multiple times?... I clicked on a video, then I read this, and then I have to click see video... I feel annoyed because there will be multiple time that I have to click.} -P5 (Male, 40s)
\end{displayquote}

For these reasons, and since only two participants drew interstitial warnings themselves, we do not evaluate interstitial warnings in our expert evaluation. It is important to note that while interstitial warnings may annoy users, they are known from prior work in other domains to be more effective than passive warnings~\cite{egelman2008you}. However, thus far, few media platforms have adopted such warnings (Twitter is a notable exception).

Similarly, participants gave mixed reviews to the non-interstitial, but the highly visible design probe shown in Figure \ref{fig:dp4} (Appendix~\ref{App:designprobes}). Participants appreciated that the design was easy to see, but found the size and text used in the warning disruptive. Experts, however, liked this design probe and suggested decreasing the thickness of the border and changing the text used (see Figure \ref{fig:1du}) to accommodate participants' preferences  (see Figure~\ref{fig:1du}). 
%
%
\section{Quantitative Evaluation: Evaluating Interpretability of Video Authenticity Indicators}
To evaluate the interpretability of the warnings created through our participatory design and expert evaluation studies, as well as to identify the most understandable set of warnings and/or warning design principles for video media provenance, we conducted a quantitative evaluation of the interpretability of a representative set of video provenance warning designs (n=1,456). 
The purpose of this work is to confirm the \textit{interpretability} of our warnings outside of our participatory design sample. However, the evaluation presented here is but a first step: just because users can understand the warnings does not necessarily mean that they will be effective at reducing the rate at which news consumers are misled by fake videos. Future work must be conducted to evaluate these warnings in the field. In this section we review the methods and results from our evaluation of these designs. All methods were approved by our institution's ethics review board.

\begin{table*}[h]
\centering
\caption{Representative Design Combinations of Icon, Signal Word, and Message Length. * indicates designs ultimately tested in our quantitative evaluation (see Section~\ref{sec:quant:methods} for more detail).}
\label{table:FinalDesigns}
\begin{tabular}{@{}rccccc@{}}
\toprule
\multicolumn{1}{c}{Design} & Signal Word & \begin{tabular}[c]{@{}c@{}}Message\\ Length\end{tabular} & Icon & Phrase Used\\ \midrule
\multicolumn{1}{l|}{CSC*} & Confirmed & Short & green \checkmark &Confirmed Source:  XYC News \\
\multicolumn{1}{l|}{CLC*} & Confirmed & Long & green \checkmark&  Confirmed Source: Video published by XYC News \\
\multicolumn{1}{l|}{CLB*} & Confirmed & Long & green border&  Confirmed Source: Video published by XYC News \\
\multicolumn{1}{l|}{UcSO*} & Unconfirmed & Short & orange !& Unconfirmed Source \\
\multicolumn{1}{l|}{UcSB*} & Unconfirmed & Short & red border & Unconfirmed Source \\
\multicolumn{1}{l|}{VSC} & Verified & Short & green \checkmark & Verified Source:  XYC News \\
\multicolumn{1}{l|}{VSB} & Verified & Short & green border & Verified Source:  XYC News \\
\multicolumn{1}{l|}{VLC*} & Verified & Long & green \checkmark & Verified Source: Video published by XYC News\\
\multicolumn{1}{l|}{UvSO*} & Unverified & Short & orange ! & Unverified Source\\
\multicolumn{1}{l|}{UvSB} & Unverified & Short & red border & Unverified Source\\ \midrule
\multicolumn{5}{l}{\textbf{Possible Survey Pairs:} CSC-UcSO\hfil CLC-UcSO\hfil CLB-UcSB\hfil VSC-UvSO\hfil VSB-UvSB\hfil VLC-UvSO}  \\ \bottomrule
\end{tabular}
\end{table*}

\begin{table*}[h!]
\centering
\caption{Quantitative Survey Results. All possible representative design variants are shown for symmetry with Table~\ref{table:FinalDesigns} but those with -- were not tested (see Section~\ref{sec:quant:methods})}.
\label{tab:finalquantrespon}
\resizebox{\textwidth}{!}{%
\begin{tabular}{@{}rlccccclllll@{}}
\toprule
\multicolumn{1}{l|}{} & \multicolumn{11}{c}{\textbf{Designs}} \\ \midrule
\multicolumn{1}{l|}{\textbf{Selection}} & \multicolumn{1}{c}{\textbf{\begin{tabular}[c]{@{}c@{}}CSC\\(N=287)\end{tabular}}} & \textbf{\begin{tabular}[c]{@{}c@{}}CLC\\(N=285)\end{tabular}} & \textbf{\begin{tabular}[c]{@{}c@{}}CLB\\(N=299)\end{tabular}} & \textbf{\begin{tabular}[c]{@{}c@{}}VSC\\
\end{tabular}} & \textbf{\begin{tabular}[c]{@{}c@{}}VSB\\
\end{tabular}} & \textbf{\begin{tabular}[c]{@{}c@{}}VLC\\(N=285)\end{tabular}} & \multicolumn{1}{c}{\textbf{\begin{tabular}[c]{@{}c@{}}UcSO\\(N=285)\end{tabular}}} & \multicolumn{1}{c}{\textbf{\begin{tabular}[c]{@{}c@{}}UcSB\\(N=299)\end{tabular}}} & \multicolumn{1}{c}{\textbf{\begin{tabular}[c]{@{}c@{}}UvSO\\(N=285)\end{tabular}}} & \multicolumn{1}{c}{\textbf{\begin{tabular}[c]{@{}c@{}}UvSB\\
\end{tabular}}} & \multicolumn{1}{c}{\textbf{\begin{tabular}[c]{@{}c@{}}FB\\(N=300)\end{tabular}}} \\ \midrule
\multicolumn{1}{c|}{1} & \multicolumn{1}{c}{26\%} & 25\% & 30\% & --
 & -- & 29\% & \multicolumn{1}{c}{34\%} & \multicolumn{1}{c}{36\%} & \multicolumn{1}{c}{34\%} &
 \multicolumn{1}{c}{--} & 
 \multicolumn{1}{c}{31\%} \\
\multicolumn{1}{c|}{2} & \multicolumn{1}{c}{49\%*} & 64\%* & 57\%* & -- 
& -- & 55\%* & \multicolumn{1}{c}{64\%*} & \multicolumn{1}{c}{57\%*} & \multicolumn{1}{c}{61\%*} &
\multicolumn{1}{c}{--} & 
\multicolumn{1}{c}{47\%} \\
\multicolumn{1}{c|}{3} & \multicolumn{1}{c}{33\%} & 38\% & 34\% & -- 
& -- & 38\% & \multicolumn{1}{c}{39\%} & \multicolumn{1}{c}{33\%} & \multicolumn{1}{c}{38\%} & \multicolumn{1}{c}{--} & 
\multicolumn{1}{c}{62\%*} \\
\multicolumn{1}{c|}{4} & \multicolumn{1}{c}{53\%} & 48\% & 36\% & -- & 
-- & 48\% & \multicolumn{1}{c}{36\%} & \multicolumn{1}{c}{33\%} & \multicolumn{1}{c}{34\%} & \multicolumn{1}{c}{--} & 
\multicolumn{1}{c}{33\%} \\
\multicolumn{1}{c|}{5} & \multicolumn{1}{c}{17\%} & 13\% & 16\% & --
& -- & 18\% & \multicolumn{1}{c}{13\%} & \multicolumn{1}{c}{15\%} & \multicolumn{1}{c}{19\%} & \multicolumn{1}{c}{--} &
\multicolumn{1}{c}{17\%} \\
\multicolumn{1}{c|}{6} & \multicolumn{1}{c}{2\%} & 1\% & 1\% & -- 
& -- & 1\% & \multicolumn{1}{c}{1\%} & \multicolumn{1}{c}{2\%} & \multicolumn{1}{c}{1\%} &
\multicolumn{1}{c}{--} & 
\multicolumn{1}{c}{2\%} \\
\multicolumn{1}{c|}{7} & \multicolumn{1}{c}{4\%} & 4\% & 4\% & -- & 
-- & 1\% & \multicolumn{1}{c}{2\%} & \multicolumn{1}{c}{4\%} & \multicolumn{1}{c}{2\%} & \multicolumn{1}{c}{--} & 
\multicolumn{1}{c}{1\%} \\ \midrule
\multicolumn{1}{c|}{OI} & \multicolumn{1}{c}{16\%} & 22\% & 24\% & -- & 
-- & 15\% & \multicolumn{1}{c}{24\%} & \multicolumn{1}{c}{23\%} & \multicolumn{1}{c}{22\%} & \multicolumn{1}{c}{--} & 
\multicolumn{1}{c}{23\%} \\ \midrule
\multicolumn{1}{l}{} &  & \multicolumn{1}{l}{} & \multicolumn{1}{l}{} & \multicolumn{1}{l}{} & \multicolumn{1}{l}{} & \multicolumn{1}{l}{} &  &  &  &  &  \\

\multicolumn{12}{l}{* = Intended answer}\\
 \multicolumn{12}{l}{1 = The location and date of the video was confirmed/verified}   \\
  \multicolumn{12}{l}{2 = The news agency that produced the video was unconfirmed/verified}  \\
\multicolumn{12}{l}{ 3 = This video provides (in) accurate information}  \\
 \multicolumn{12}{l}{4 = XYC News is (not) a trustworthy  source}   \\
  \multicolumn{12}{l}{5 = It is recommended that I (do not) watch this video}   \\
  \multicolumn{12}{l}{6 = Other}  \\
\multicolumn{12}{l}{7 = I don't know}   \\ 
\multicolumn{12}{l}{OI = Participants that only selected the intended interpretation}\\
\end{tabular}%
}
\end{table*}

\subsection{Methods}
\label{sec:quant:methods}
We quantitatively evaluate a representative set of the designs most preferred by both participants and experts. Specifically, as most participant designs relied heavily on check mark / exclamation point icons, and phrases underneath the video frame, we test warnings that include these elements. As part of these designs, we test the participant-preferred phrase \textit{(un)verified} against the expert-preferred phrase \textit{(un)confirmed}. Additionally, since experts worried that such designs would be insufficient to draw participant's attention and thus suggested using a more visible warning design with a border (see Figure \ref{fig:1du}), we evaluate designs that contain borders against those that use the participant-preferred check mark / exclamation point icons. Finally, we also evaluate one state-of-the-art misinformation warning message, used by Facebook~\cite{oldfb2} (see Figure \ref{fig:fbwarning} in Appendix~\ref{App:designprobes}), as a baseline for interpreting our results.\footnote{We use Facebook's 2017 warning design as a baseline because its goal -- to raise awareness of potentially suspicious information -- is closest to the goal of our work. Facebook's current warning design suggests related articles for user education, rather than aiming to provide users with signal about the present content~\cite{oldfbblob}.} 

Each design we test is comprised of a signal word, icon, and a phrase as shown in Table~\ref{table:FinalDesigns}. To avoid excessive use of resources, we do not evaluate every single possible combination of signal word, icon and phrase shown in Table~\ref{table:FinalDesigns}. Instead, we evaluate design attributes sequentially. We first compare signal words in designs that were otherwise identical (we used long, green icon warnings for this comparison as they are most reflective of participant designs); next, we compared length for the most effective signal word; and finally we compared icons for the most effective signal words and lengths. See Appendix \ref{App:DesignImagesinSurvey} for the designs ultimately tested and Table~\ref{tab:finalquantrespon} for the results of these tests.

\subsubsection{Survey Methodology} 
In August 2020, we tested these designs by distributing five surveys
using MTurk~\footnote{See Section~\ref{sec:limitations} for a discussion of the use of MTurk.}. Four of the surveys tested a pair of designs: one that indicated a positively authenticated media instance and one that indicated a non-authenticated media instance; the baseline survey tested only Facebook's warning, which provides no notification for ``true'' content, only a warning for potentially false content. The surveys included a digital consent form, design pair interpretation questions, and demographic questions. The warnings in each pair were displayed randomly in the surveys to eliminate order effects. The design interpretation questions were multiple selection and asked participants to select the options that best fit their interpretation of each warning they were shown (see Table~\ref{tab:finalquantrespon} for interpretation options). Based on a power analysis, we aimed to recruit at least 250 participants per survey, in order to observe small- to medium-sized differences in the interpretability of the designs.

 Each survey took between two and three minutes to complete and participants were compensated \$.40 to complete each one. In order to participate, participants had to be in the United States of America, have a 95\% or higher approval rating~
\cite{peer2014reputation}, and could only complete one of the multiple surveys we posted. 

\subsubsection{Participants} 
In total, 1,456 people participated in this study. Additional demographic information can be found in Table~\ref{table:demoquant} in Appendix~\ref{App:DesignImagesinSurvey}. 

\subsubsection{Analysis}
 Using a Chi-square test for homogeneity, we compared the proportion of participants that made a particular selection for a design to another design. As comparisons were limited and planned, we did not apply correction~\cite{armstrong2014use}. All calculations were completed using R, an open-source software for statistical computing~\cite{Rstuff}. 

\subsection{Results}
The purpose of each design is to notify users that the video source was authenticated or that it was not. Overall, we find that when the message informed the user that the video source was authenticated, a larger proportion of participants were able to correctly interpret the message as intended when the message included the signal word "confirmed", a longer more specific phrase describing what was confirmed (the source of the video), and a green checkmark (CLC 64\%) or green border (CLB 57\%) (see Table~\ref{tab:finalquantrespon}). When the source was not authenticated, more participants responded correctly to the signal word unconfirmed than unverified and responded correctly equally as often when this signal word was paired with an exclamation point icon (UcSO 64\%) or a red border (UcSB 57\%). 

\subsubsection{Effect of Individual Design Choices}
 When participants were asked to interpret the meaning of the warning design they viewed, they were asked to select all of the options that applied. We tested the significant proportional differences between designs to see how changes in design impact interpretation. In order to do this, we compared warnings that differed in one design choice - phrase length, signal word, and icon. Below, we report on the effect of each design choice.

  We first evaluated the effect of \textbf{signal word} (e.g., verified versus confirmed) on warnings that were otherwise identical. Comparing VLC to CLC, we find that a larger proportion of participants selected the intended interpretation ($p$ = 0.03) when shown the warning with the ``confirmed'' phrase (CLC, 64\%) than with the ``verified'' phrase (VLC, 55\%). 
  
  When examining whether users \textit{only} select the intended interpretation of the warning, we observe that only 22\% of respondents selected only the intended interpretation for CLC and 15\% did so for VLC (a significant difference with $p$ = 0.031). This suggests that such warnings are highly likely to be overgeneralized (see Section~\ref{sec:quant:results:overgeneralize} for further discussion); however, the ``confirmed'' phrasing still outperforms the ``verified'' phrasing, confirming the concerns our experts raised about overgeneralization from the ``verified'' term. 
  
Next, we evaluated the impact of \textbf{phrase length} using the ``confirmed'' keyword. When comparing CSC vs CLC, the longer phrase resulted in an increase in the proportion of participants that selected the intended interpretation (49\% with CSC vs. 64\% with CLC, $p$ = 0.001). Phrase length did not have a significant impact on the degree of overgeneralization (e.g., the proportion of respondents who selected only the intended interpretation).

Finally, we compared the use of \textbf{icons} to colored borders by comparing CLC to CLB and UcSO to UcSB. While we observe no significant difference in the proportion of respondents correctly interpreting CLC and CLB, we do observe that the use of a border leads to better specificity in warning interpretation: a significantly higher proportion of participants ($p$ = 0.005) interpreted the warning CLC warning to also mean that XYC news, the confirmed producer of the video, was a trustworthy source. Thus, it appears that the border design may better focus participants' interpretations to the specific content -- the video -- highlighted by the border. We find no significant differences in users' interpretations of the unconfirmed warnings with a red border vs. with an exclamation point. 

 \subsubsection{Performance Against the Baseline \& Overgeneralization}
 \label{sec:quant:results:overgeneralize}
To evaluate our warnings against the current state-of-the-art misinformation warnings for text content, we evaluate the interpretability of our best warning designs against Facebook's previously used misinformation warning (Table \ref{tab:finalquantrespon}). We find that our warnings equally as interpretable  as the existing state-of-the-art (64\% of respondents selecting intended interpretation for UcSO vs. 62\% for FB $p$ > 0.05; 57\% for UcSB vs. 62\% for FB $p$ > 0.05). Further, we find that participants did not overgeneralize the meaning of the user-created warnings  significantly different than they do Facebook warnings (24\% of respondents select only the intended interpretation for UcSO vs. 23\% for FB, $p>0.05$; 23\% UcSB vs. 23\% for FB, $p$ > 0.05). 

\section{Limitations}
\label{sec:limitations}
The qualitative portions of our work (Sections~\ref{sec:interview} and~\ref{sec:designstudy}) have limitations common to qualitative research. First, the generalizability of our results is limited by the constraint of the qualitative sample size. We took care to recruit a sociodemographically diverse sample to increase the diversity of experiences captured in our work, but it is possible that we may have omitted an important dimension of diversity in our recruitment and thus not included a full spectrum of participant experiences. To partially counteract this risk, we did conduct follow up quantitative studies both to determine the prevalence of certain key interview results (e.g., Americans' awareness of deep fakes) and to validate that the designs produced through our qualitative work were understandable to  larger groups of participants. Second, interviews rely on participants' comfort with sharing information with the interviewer and their ability answer to the interviewer's questions. It is possible that despite our best efforts to follow best practice in our interviews and participatory design sessions, participants did not fully share their experiences in response to our questions.

The quantitative portions of our work are limited by the use of MTurk for recruiting our survey participants. Further, although we added questions to detect bots and limited participants to one of the seven surveys, we recognize that there are ways to circumvent these measures, and our results could include response from bots or repeat survey participants. Although Amazon Mechanical Turk (MTurk) workers are different from the general population in many ways, prior work has found that, for security and privacy related issues, MTurk workers are as representative of the experiences of Americans under 50 as opt-in survey panels with census-quota determined demographics~\cite{redmiles2019well}. Given this context, and that we took great care to obtain diverse perspectives in our qualitative work, we proceeded with using MTurk samples for our quantitative confirmations. However, our results should be interpreted with the appropriate sample limitations in mind.

Finally, as aforementioned, our quantitative evaluation evaluates the interpretability of our warnings but does not evaluate the impact of our warnings on user behavior. Future work should seek to evaluate video provenance warnings in the wild, to understand their mitigation effect for fake video media.
\section{Discussion}
Our interview results confirm the findings of prior work \cite{zaryan2017truth,fogg2003users} that users have multiple heuristics for combating misinformation that have been developed over time and exposure to such content. However, while our participants appeared confident and were quick to discuss fact-checking methods for text-based news articles, many were less sure of themselves and lacked heuristics for detecting misinformation depicted in images and videos. In sum, our interview results suggest that there are significant differences in how people experience and respond to text-, image-, and video-based misinformation. Thus, future work on misinformation and its warnings should explicitly consider the media's type.

More specifically, some participants simply expressed that they did not have a method for evaluating image or video content, while others mentioned that ``that they would just know'' if a video or image had been edited. However, as video editing techniques improve, it will increasingly become more difficult for the untrained eye to detect edited or deep fakes videos, especially when they are presented as real from a trusted source. Due to this, participants expressed concern about not having the necessary context to evaluate videos and images; with some participants even assuming all videos and images are fake unless there is relevant text to provide the context. Our design work takes a first step toward filling this gap, exploring best design practices for the creation of warnings that can provide the user with additional context regarding the authenticity of video media. 

In line with prior work \cite{metzger2010social, fogg2003users, lee2013tweet}, we find that source is a particularly critical heuristic for identifying misinformation, regardless of media mode. Thus, providing warnings that leverage users' existing emphasis on source by providing information on media provenance -- whether the media they are seeing is actually unmodified content from the claimed source -- can be the first step in the right direction for combating misinformation, especially for video content. 

Such warnings must balance avoiding misleading users and reducing user engagement with content by frustrating them with warnings. Our design study results suggest that such warnings should avoid interrupting users' consumption flow -- for example, by being brief, easy to interpret, and passive rather than interstitial -- and should be highly specific (as also suggested by prior work~\cite{egelman2008you, ross2018fake, wogalter2002based}). 

Our quantitative evaluation finds that the most interpretable video provenance warnings are those that use the word \textit{confirm}, along with specific details (such as what the confirmed source of the video is). We found that either a red exclamation point or a red border for unconfirmed videos and a green check mark (for confirmed videos) or a green border around the video were equally interpretable, however the use of the green check mark may be overgeneralized as indicating that the news source is trustworthy as well.

\textbf{Best practices for future misinformation design work.} We leveraged both participatory design with end-users and expert feedback to develop our designs. We find that while participants' designs reflected one of the three most effective design elements -- longer and more specific messages, it was the experts who suggested the other two effective elements: the use of a more specific signal word (confirmed vs. verified) and the border design; although the ultimate design of the latter was informed by participants' feedback as well. Thus, our results suggest that future work seeking to design such warnings should conduct \textit{both} participatory design with end-users and expert evaluations to develop the most effective warnings. 

\textbf{Applicability of designs beyond video provenance.} In this work, we focus on designs which indicate the authenticity of media generated by a provenance system. When media content can be cryptographically authenticated back to its publisher, these provenance indicators provide a secure guarantee that the media is authentic. On the other hand, automated fake media detection systems produce a score, and possibly a confidence level, that can be used to indicate the probability that the media is ``fake''. Our study results can also be applied to fake media detection to some extent, but additional research is required to understand the best method to convey to users the notion of probabilistic detection of manipulated media -- rather than the binary and definitive notion of authenticated vs. non-authenticated addressed in this work.

\textbf{Watch out for misinformation warning overgeneralization.} Finally, our findings highlight a broad concern for misinformation warnings that address any mode of media: overgeneralization. Both our misinformation warnings and the baseline state-of-the-art warning from Facebook, against which we evaluated our warnings, 
exhibited high degrees of overgeneralization: while a majority of respondents selected the intended interpretation of the warning, among other interpretations, just under a quarter selected only the intended warning interpretation. In addition to the intended interpretation of our warnings, users also took the warnings to mean that they should not watch the video because, for example, the content was inaccurate or the source was untrustworthy.  

Thus, future work may seek to critically examine the consequences of overgeneralization of misinformation warnings. Evaluations of such warnings should take care to evaluate this property, and warning designers should carefully consider the impact they want to have with the warnings they provide given users tendency to generalize past the intended effect of the warning. 
\section{Conclusion}
In this work, we conducted a four-part, mixed methods study to investigate Americans' experiences with video misinformation and design warnings to counteract such content. We conducted an interview study with 24 diverse Americans, identifying that the availability of heuristics through which users can identify misinformation and their confidence in applying those heuristics varies by media mode (text vs. image vs. video) and identifying that trust in the source of the information is a unifying misinformation heuristic across media modes. 

Building on users' existing use of information source and the emerging viability of technical systems for verifying the provenance (authenticity) of video source as a promising way of combating fake media content, we develop warnings and design guidelines for communicating the provenance of video content to end-users. To this end, we conducted participatory design studies with 19 diverse Americans to develop and qualitatively evaluate designs for warnings to alert users to the authenticity of video media content. We further evaluated these warnings through expert design reviews and larger-scale quantitative evaluations with crowd workers. We find that a combination of user and expert insights are necessary to define the most interpretable warnings, and raise concerns around the potential for users to overgeneralize misinformation warnings regarding video or even text information. Our results offer concrete insights for misinformation warning designers and suggest multiple directions for future research.

\bibliographystyle{IEEEtran}
\bibliography{main}


\clearpage
\appendices
\section{Interview Demographics}
\label{App:InterDemo}
In this appendix, we describe the demographics of the participants of the two parts of our interview study.  Table \ref{table:DemoInterview} provides the demographics of the interview participants. The demographics of the MTurk participants who answered questions regarding their knowledge of deep fake videos and reverse image search via a survey are displayed in Table~\ref{table:DemoDeepSurvey}.\\
\begin{table} [h!]
\caption{Demographics for Interview Participants}
\label{table:DemoInterview}
\centering
\begin{tabular}{@{}rcc@{}}
\toprule
\multicolumn{1}{c|}{\textbf{Demographics}} & \multicolumn{2}{c}{\textbf{Participants}} \\ \midrule
\multicolumn{1}{l}{ (N=24)} & \multicolumn{1}{l}{n} & \multicolumn{1}{l}{\%} \\ \midrule
\multicolumn{3}{l}{\textbf{Gender }}                                                                                                                     \\ 
\midrule
\multicolumn{1}{r|}{Female}  & 13   & 54\%                                     \\
\multicolumn{1}{r|}{Male}     & 11                                 & 46\%                                     \\ 
\multicolumn{1}{r|}{Non-binary} & 0    & 0\%   \\
\multicolumn{1}{r|}{Prefer not to answer} & 0    & 0\%   \\
\midrule
\multicolumn{3}{l}{\textbf{Age }}                                                                                                                        \\ 
\midrule
\multicolumn{1}{r|}{18-30 }    & 4                                  & 17\%                                     \\
\multicolumn{1}{r|}{31-40}  & 9                                 & 38\%                                     \\
\multicolumn{1}{r|}{41-50}  & 5                               & 21\%                                     \\
\multicolumn{1}{r|}{Over 50}  & 6                                  & 25\%                                     \\ 
\multicolumn{1}{r|}{Prefer not to answer} & 0    & 0\%   \\
\midrule
\multicolumn{3}{l}{\textbf{Race }}                                                                                                                       \\ 
\midrule
\multicolumn{1}{r|}{White}  & 12  & 50\%                                     \\
\multicolumn{1}{r|}{Hispanic or Latino}  & 1    & 4\%                                      \\
\multicolumn{1}{r|}{Black or African American}  & 3      & 13\%                                     \\
\multicolumn{1}{r|}{Native American}   & 1    & 4\%                                      \\
\multicolumn{1}{r|}{Asian }   & 5   & 21\%  \\
\multicolumn{1}{r|}{Other}   & 2 & 8\%                                      \\ 
\multicolumn{1}{r|}{Prefer not to answer} & 0    & 0\%   \\
\midrule
\multicolumn{3}{l}{\textbf{Education }}                                                                                                                  \\ 
\midrule
\multicolumn{1}{r|}{No college}  & 1  & 4\%                                      \\
\multicolumn{1}{r|}{Some college but no degree } & 5   & 21\%                                     \\
\multicolumn{1}{r|}{Associate's degree} & 5   & 21\%                                     \\
\multicolumn{1}{r|}{Bachelor's degree }  & 9  & 38\%                                     \\
\multicolumn{1}{r|}{Advanced degree} & 4   & 17\%                                     \\ 
\multicolumn{1}{r|}{Prefer not to answer} & 0    & 0\%   \\
\midrule
\multicolumn{3}{l}{\textbf{Employment }}                                                                                                                 \\ 
\midrule
\multicolumn{1}{r|}{Employed}  & 11   & 46\%                                     \\
\multicolumn{1}{r|}{Self-employed} & 4    & 17\%                                     \\
\multicolumn{1}{r|}{Not working}   & 5  & 21\%                                      \\
\multicolumn{1}{r|}{A homemaker} & 3  & 13\%                                     \\
\multicolumn{1}{r|}{A student}   & 1    & 4\%                                      \\
\multicolumn{1}{r|}{Prefer not to answer} & 0    & 0\%   \\
\midrule
\multicolumn{3}{l}{\textbf{Income }}                                                                                                                     \\ 
\midrule
\multicolumn{1}{r|}{Less than \$30,000}  & 4    & 17\%                                     \\
\multicolumn{1}{r|}{\$30,000 to under \$65,000}  & 6 & 25\%                                     \\
\multicolumn{1}{r|}{\$65,000 or more} & 13  & 54\%                                     \\
\multicolumn{1}{r|}{Prefer not to answer} & 1    & 4\%  \\
\bottomrule
\end{tabular}
\end{table}

\begin{table}[]
\caption{Deep Fake Survey Demographics}
\label{table:DemoDeepSurvey}
\centering
\begin{tabular}{@{}rcc@{}}
\toprule
\multicolumn{1}{c|}{\textbf{Demographics}} & \multicolumn{2}{c}{\textbf{Participants}} \\ \midrule
\multicolumn{1}{l}{ (N=3,552)} & \multicolumn{1}{l}{n} & \multicolumn{1}{l}{\%} \\ \midrule
\multicolumn{1}{l}{\textbf{Gender}} & \multicolumn{1}{l}{} & \multicolumn{1}{l}{} \\ \midrule
\multicolumn{1}{r|}{Female} & 1488 & 42\% \\
\multicolumn{1}{r|}{Male} & 2048 & 58\% \\
\multicolumn{1}{r|}{Non-binary} & 10 & 0\% \\
\multicolumn{1}{r|}{Prefer not to answer} & 6 & 0\% \\ \midrule
\multicolumn{1}{l}{\textbf{Age}} & \multicolumn{1}{l}{} & \multicolumn{1}{l}{} \\ \midrule
\multicolumn{1}{r|}{18-30} & 1229 & 35\% \\
\multicolumn{1}{r|}{31-40} & 1182 & 33\% \\
\multicolumn{1}{r|}{41-50} & 632 & 18\% \\
\multicolumn{1}{r|}{Over 50} & 498 & 14\% \\
\multicolumn{1}{r|}{Prefer not to answer} & 11 & 0\% \\ \midrule
\multicolumn{1}{l}{\textbf{Race}} & \multicolumn{1}{l}{} & \multicolumn{1}{l}{} \\ \midrule
\multicolumn{1}{r|}{White} & 2447 & 69\% \\
\multicolumn{1}{r|}{Hispanic or Latino} & 160 & 5\% \\
\multicolumn{1}{r|}{Black or African American} & 706 & 20\% \\
\multicolumn{1}{r|}{Native American or American Indian} & 99 & 3\% \\
\multicolumn{1}{r|}{Asian} & 231 & 7\% \\
\multicolumn{1}{r|}{Native Hawaiian or Pacific Islander} & 5 & 0\% \\
\multicolumn{1}{r|}{Other} & 8 & 0\% \\
\multicolumn{1}{r|}{Prefer not to answer} & 11 & 0\% \\ \midrule
\multicolumn{1}{l}{\textbf{Education}} & \multicolumn{1}{l}{} & \multicolumn{1}{l}{} \\ \midrule
\multicolumn{1}{r|}{No college experience} & 189 & 5\% \\
\multicolumn{1}{r|}{Some college but no degree} & 307 & 9\% \\
\multicolumn{1}{r|}{Associate's degree} & 228 & 6\% \\
\multicolumn{1}{r|}{Bachelor's degree} & 2153 & 61\% \\
\multicolumn{1}{r|}{Advanced degree} & 666 & 19\% \\
\multicolumn{1}{r|}{Prefer not to answer} & 9 & 0\% \\ \midrule
\multicolumn{2}{l}{\textbf{Employment}} & \multicolumn{1}{l}{} \\ \midrule
\multicolumn{1}{r|}{Employed} & 2622 & 74\% \\
\multicolumn{1}{r|}{Self-employed} & 528 & 15\% \\
\multicolumn{1}{r|}{Out of work} & 204 & 6\% \\
\multicolumn{1}{r|}{A homemaker} & 87 & 2\% \\
\multicolumn{1}{r|}{A student} & 66 & 2\% \\
\multicolumn{1}{r|}{Prefer not to answer} & 22 & 1\% \\ \midrule
\multicolumn{1}{l}{\textbf{Income}} & \multicolumn{1}{l}{} & \multicolumn{1}{l}{} \\ \midrule
\multicolumn{1}{r|}{Less than \$30,000} & 791 & 22\% \\
\multicolumn{1}{r|}{\$30,000 to under \$40,000} & 433 & 12\% \\
\multicolumn{1}{r|}{\$40,000 to under \$50,000} & 598 & 17\% \\
\multicolumn{1}{r|}{\$50,000 to under \$80,000} & 1066 & 30\% \\
\multicolumn{1}{r|}{\$80,000 or more} & 664 & 19\% \\
\multicolumn{1}{r|}{Prefer not to answer} & 42 & 1\% \\
\bottomrule
\end{tabular}
\end{table}

\clearpage

\section{Participatory Design Probes}
\label{App:designprobes}
\begin{figure*}[ht!]
    \centering
\begin{subfigure}{0.45\textwidth}
 \centering
  \includegraphics[scale=.45]{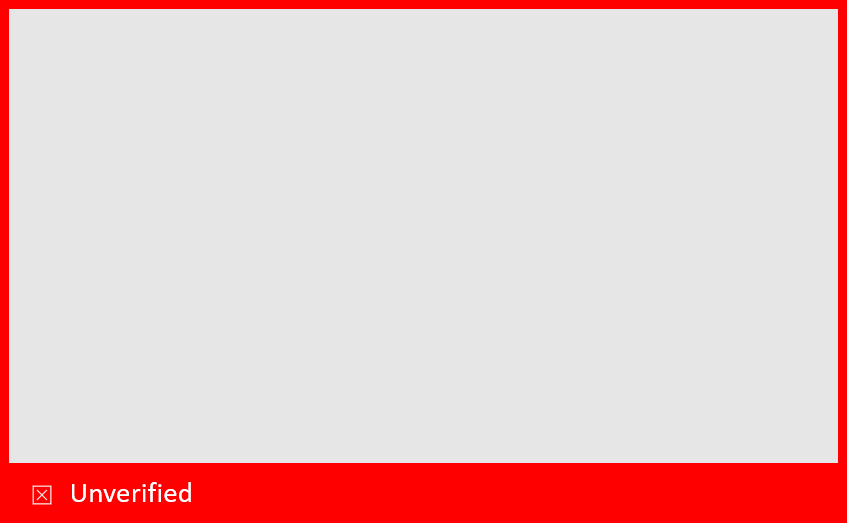}
  \caption{}
  \label{fig:dp1}
\end{subfigure}\hfil
\begin{subfigure}{.45\textwidth}
   \centering
  \includegraphics[scale =.45]{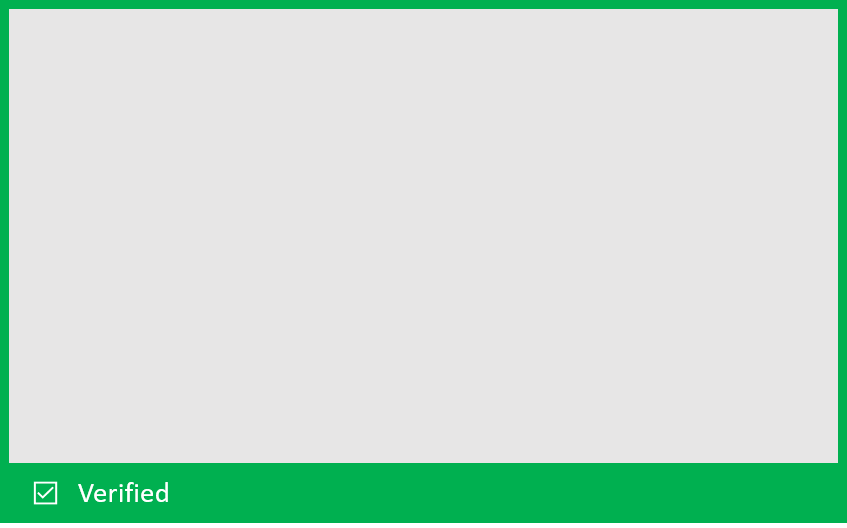}
  \caption{}
  \label{fig:dp2}
\end{subfigure}

\medskip
\begin{subfigure}{0.45\textwidth}
 \centering
 \includegraphics[scale =.4]{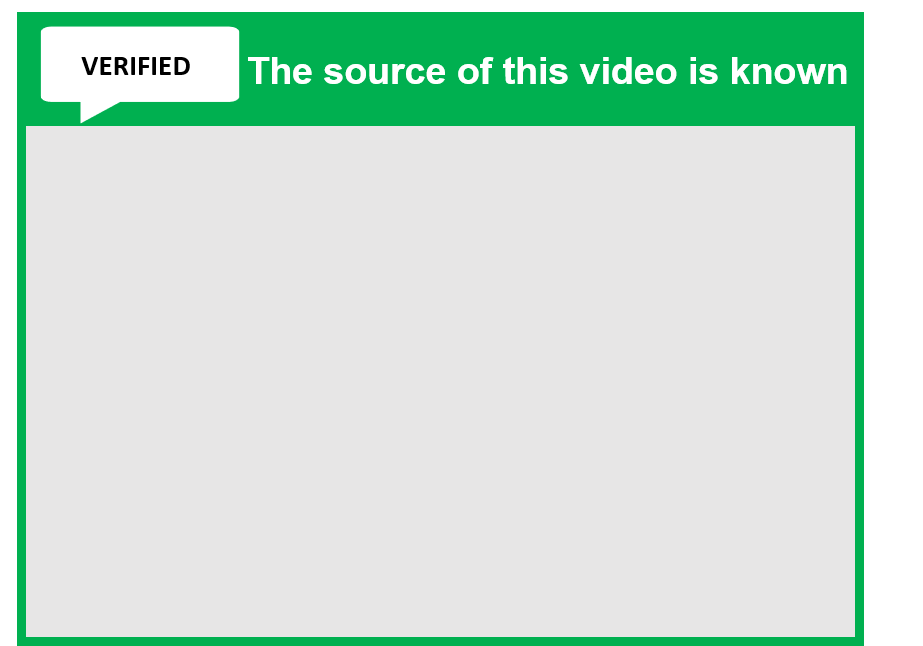}
  \caption{}
  \label{fig:dp3}
\end{subfigure}\hfil
\begin{subfigure}{.45\textwidth}
   \centering
  \includegraphics[scale =.4]{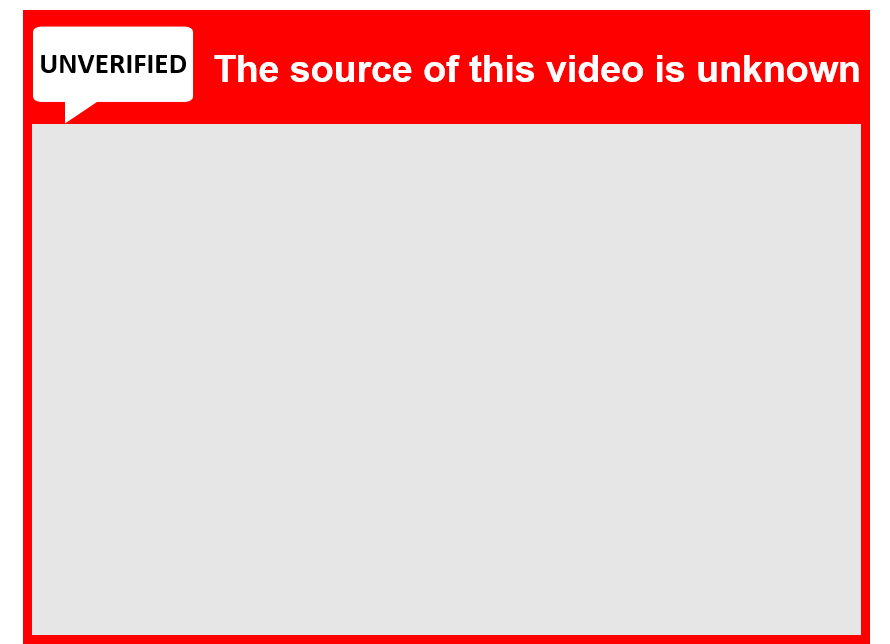}
  \caption{}
  \label{fig:dp4}
\end{subfigure}
\caption{Warning designs inspired by Slate's fake news detector browser extension \cite{slate}}
\label{fig:dpborder}
\end{figure*}

\begin{figure*}
\centering
\begin{subfigure}{0.45\textwidth}
 \centering
  \includegraphics[scale=.6]{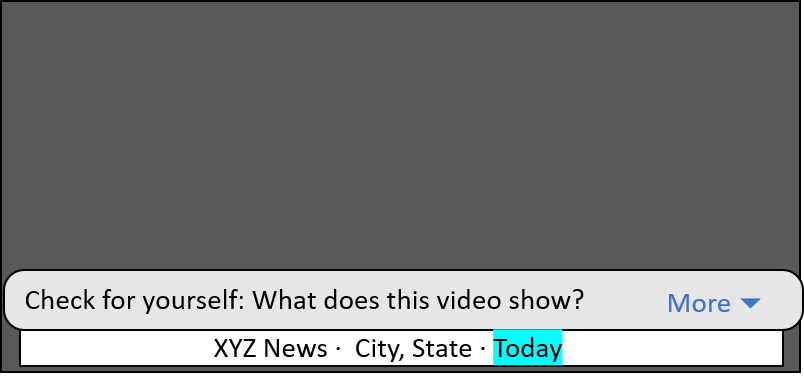}
  \caption{}
  \label{fig:dp5}
\end{subfigure}

\medskip
\begin{subfigure}{.45\textwidth}
   \centering
  \includegraphics[scale =.6]{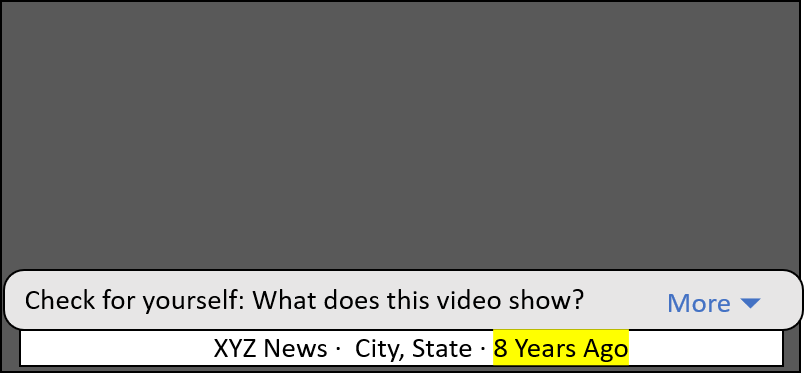}
  \caption{}
  \label{fig:dp6}
\end{subfigure}

\medskip
\begin{subfigure}{0.45\textwidth}
 \centering
  \includegraphics[scale=.5]{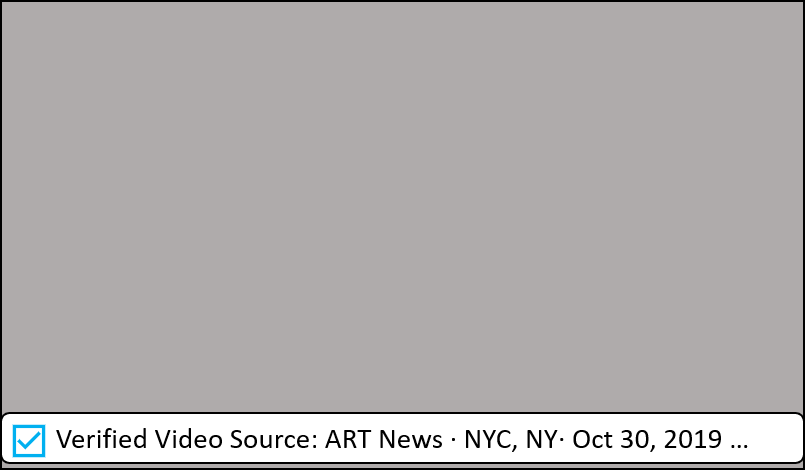}
  \caption{}
  \label{fig:dp7}
\end{subfigure}

\medskip
\begin{subfigure}{.45\textwidth}
   \centering
    \includegraphics[scale=.5]{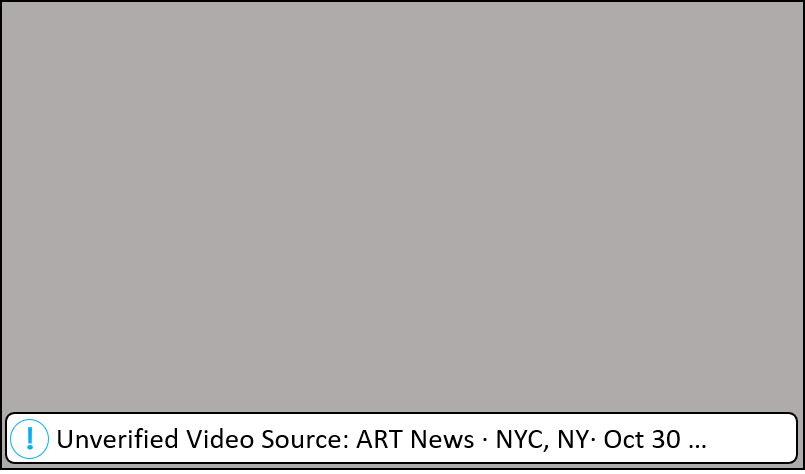}
  \caption{}
  \label{fig:dp8}
\end{subfigure}
\caption{Warning designs inspired by the News Provenance Project \cite{koren2019introducing}}
  \label{fig:dpNPP}
\end{figure*}

\begin{figure*}
\begin{subfigure}{0.45\textwidth}
   \centering
  \includegraphics[scale =.75]{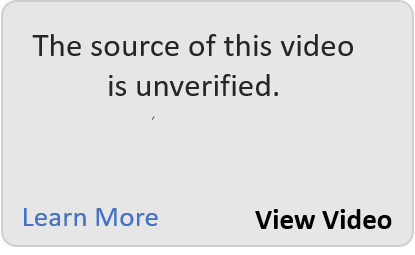}
  \caption{}
  \label{fig:dp9}
\end{subfigure}\hfil
\begin{subfigure}{0.45\textwidth}
 \includegraphics[scale =.75]{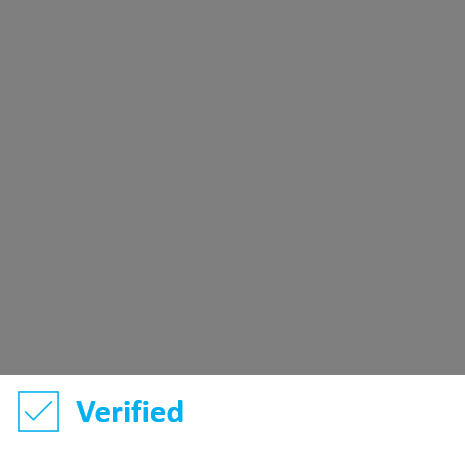}
  \caption{}
  \label{fig:dp10}
\end{subfigure}

\medskip
\begin{subfigure}{0.45\textwidth}
   \centering
  \includegraphics[scale =.75]{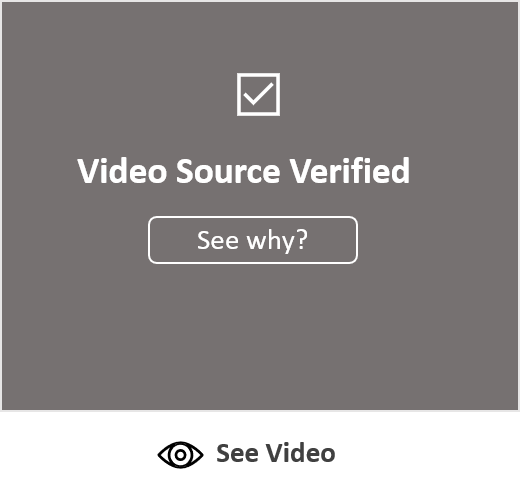}
  \caption{}
  \label{fig:dp11}
\end{subfigure}\hfil
\begin{subfigure}{0.45\textwidth}
  \includegraphics[scale =.75]{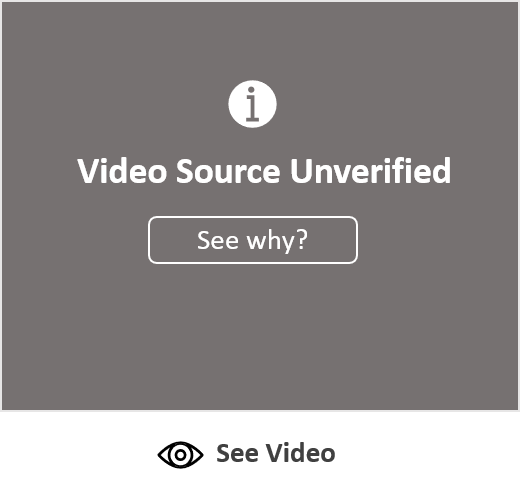}
  \caption{}
  \label{fig:dp12}
\end{subfigure}
 \caption{Warning design inspired by current Twitter warnings \cite{twitter}}
  \label{fig:dptwitter}
\end{figure*}

\begin{figure*}
\begin{subfigure}{0.45\textwidth}
 \centering
  \includegraphics[scale =.75]{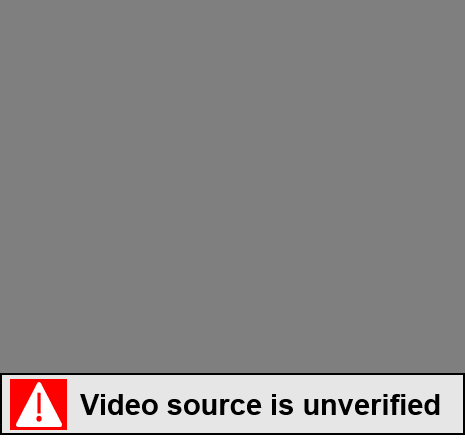}
  \caption{Warning design inspired by Facebook warning discontinued in 2017 \cite{oldfb2}}
  \label{fig:dp13}
\end{subfigure}\hfil
\begin{subfigure}{.45\textwidth}
   \centering
  \includegraphics[scale =.75]{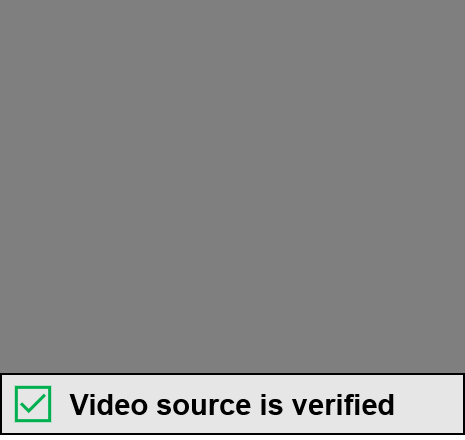}
  \caption{}
  \label{fig:dp14}
\end{subfigure}

\medskip
\begin{subfigure}{0.45\textwidth}
 \centering
  \includegraphics[scale=.75]{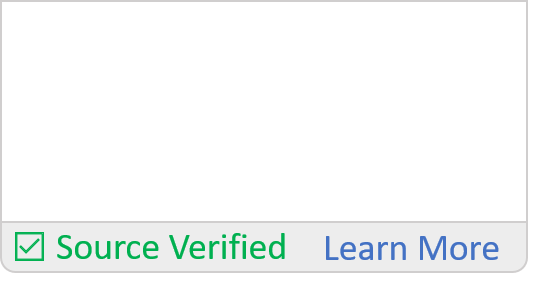}
  \caption{}
  \label{fig:dp15}
\end{subfigure}\hfil
\begin{subfigure}{.45\textwidth}
   \centering
  \includegraphics[scale =.75]{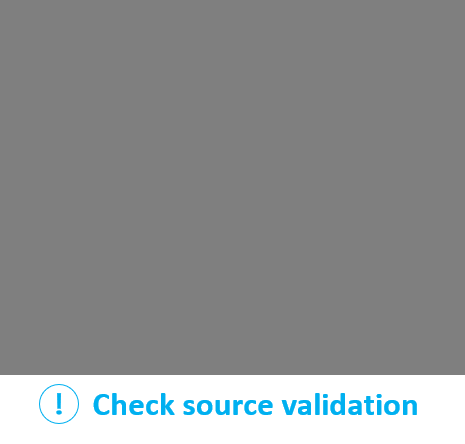}
  \caption{}
  \label{fig:dp16}
\end{subfigure}
\caption{This figure includes four warning design probes where the warning information is below the video}
\label{fig:Designprobes}
\end{figure*}

\begin{figure*} [ht!]
    \centering
   \begin{subfigure}{0.45\textwidth}
 \centering
  \includegraphics[scale=.45]{figures/pd1a.png}
\end{subfigure}\hfil
\begin{subfigure}{.45\textwidth}
   \centering
  \includegraphics[scale =.45]{figures/pd767.png}
\end{subfigure}

\medskip
 \centering
   \begin{subfigure}{0.45\textwidth}
 \centering
  \includegraphics[scale=.45]{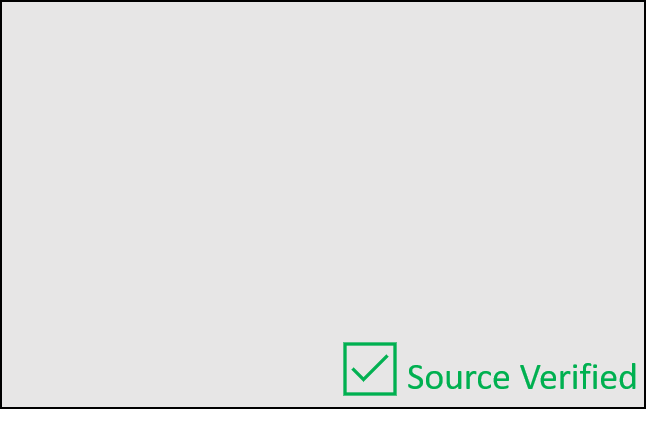}
\end{subfigure}\hfil
\begin{subfigure}{.45\textwidth}
   \centering
  \includegraphics[scale =.45]{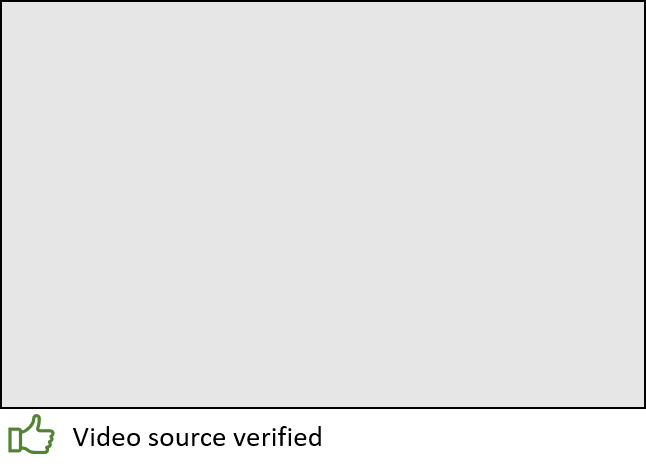}
\end{subfigure}

\medskip
 \centering
   \begin{subfigure}{0.45\textwidth}
 \centering
  \includegraphics[scale=.45]{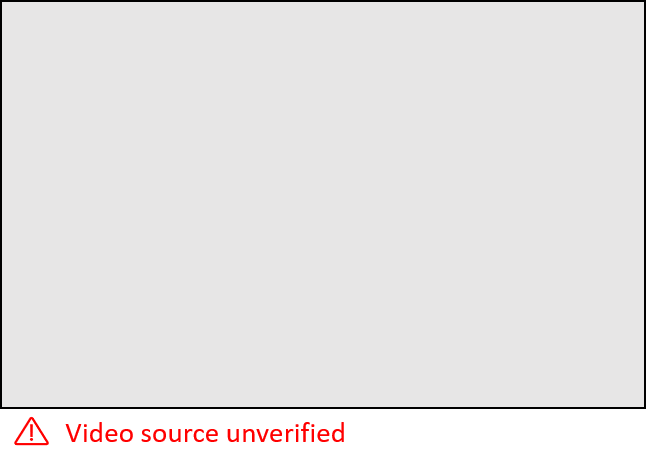}
\end{subfigure}\hfil
\begin{subfigure}{.45\textwidth}
   \centering
  \includegraphics[scale =.45]{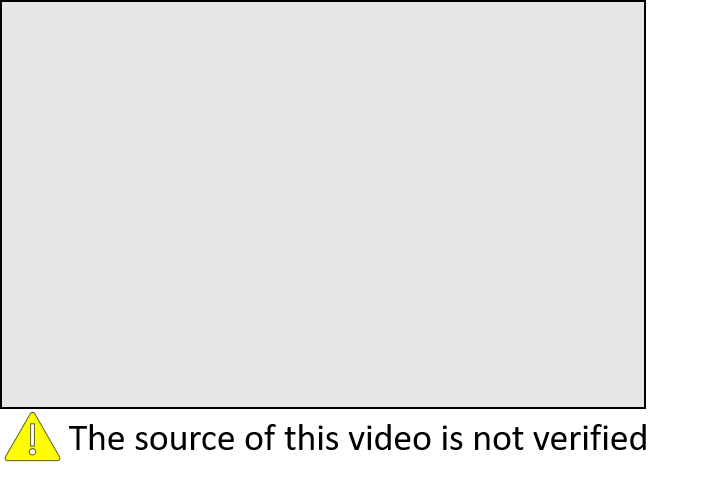}
\end{subfigure}

\medskip
 \centering
   \begin{subfigure}{0.45\textwidth}
 \centering
  \includegraphics[scale=.45]{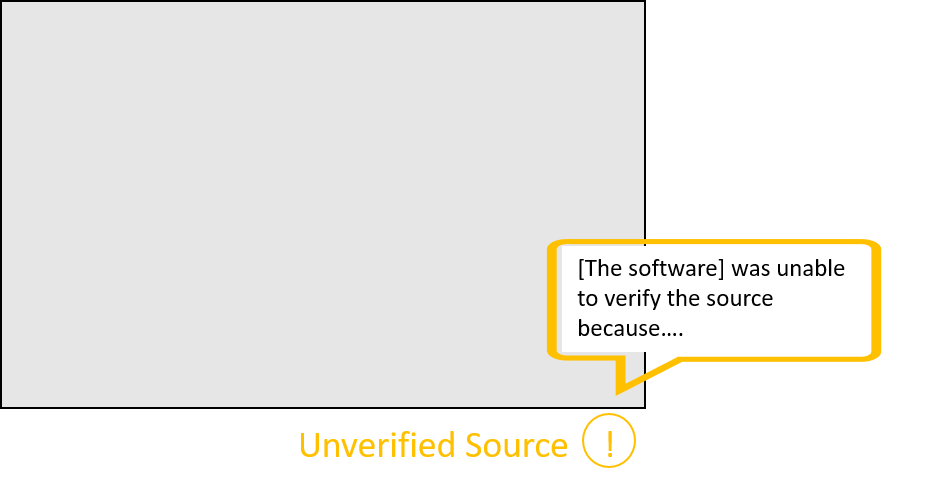}
\end{subfigure}\hfil
\begin{subfigure}{.45\textwidth}
   \centering
  \includegraphics[scale =.45]{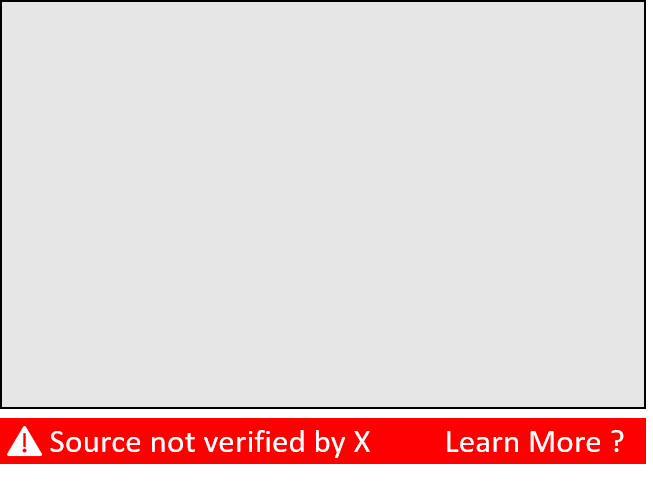}
\end{subfigure}
    \caption{Subset of Participant Warning Designs}
    \label{fig:partdesign2}
\end{figure*}

 This appendix provides the design probes used in the participatory design study. Figures~\ref{fig:dpborder},~\ref{fig:dpNPP},~\ref{fig:dptwitter}, and~\ref{fig:Designprobes} depict the screenshots of these design probes. Figure~\ref{fig:partdesign2} displays a subset of the notification designs created by participants. Finally, Table~\ref{table:DemoPartDesign} displays the demographics of the participants from this study.\\
 
\begin{table}[ht]
\caption{Demographics for Participatory Design Study}
\label{table:DemoPartDesign}
\centering
\begin{tabular}{@{}rcc@{}}
\toprule
\multicolumn{1}{c|}{\textbf{Demographics}} & \multicolumn{2}{c}{\textbf{Participants}} \\ \midrule
\multicolumn{1}{l}{ (N=19)} & \multicolumn{1}{l}{n} & \multicolumn{1}{l}{\%} \\ 
\midrule
\multicolumn{3}{l}{\textbf{Gender} }                                             \\ 
\midrule
\multicolumn{1}{r|}{Female}                                  & 10             & 53\%                 \\
\multicolumn{1}{r|}{Male}                                  & 9              & 47\%                 \\ 
\multicolumn{1}{r|}{Non-binary} & 0 & 0\% \\
\multicolumn{1}{r|}{Prefer not to answer} & 0 & 0\% \\ \midrule
\multicolumn{3}{l}{\textbf{Age} }                                                \\ 
\midrule
\multicolumn{1}{r|}{18-30}                                    & 5              & 26\%                 \\
\multicolumn{1}{r|}{31-40}                                    & 7              & 37\%                 \\
\multicolumn{1}{r|}{41-50}                                    & 1              & 5\%                 \\
\multicolumn{1}{r|}{Over 50}                                 & 6              & 32\%                 \\ 
\midrule
\multicolumn{3}{l}{\textbf{Education} }                                          \\ 
\midrule
\multicolumn{1}{r|}{No college}               & 0          & 0\%                 \\
\multicolumn{1}{r|}{Some college but no degree}               & 4              & 21\%                 \\
\multicolumn{1}{r|}{Associate's degree }                     & 2              & 11\%                 \\
\multicolumn{1}{r|}{Bachelor's degree}                        & 10             & 53\%                 \\
\multicolumn{1}{r|}{Advanced degree}                        & 3              & 16\%                 \\ 
\midrule
\multicolumn{3}{l}{\textbf{Race} }                                               \\ 
\midrule
\multicolumn{1}{r|}{White}                                    & 7              & 37\%                 \\
\multicolumn{1}{r|}{Hispanic or Latino}                       & 1              & 5\%                  \\
\multicolumn{1}{r|}{Black or African American}              & 5              & 26\%                 \\
\multicolumn{1}{r|}{Native American}      & 1              & 5\%                  \\
\multicolumn{1}{r|}{Asian}                                    & 4              & 21\%                 \\
\multicolumn{1}{r|}{Other}                                    & 2              & 11\%                 \\ 
\multicolumn{1}{r|}{Prefer not to answer} & 0 & 0\% \\
\midrule
\multicolumn{3}{l}{\textbf{Employment} }                                         \\ 
\midrule
\multicolumn{1}{r|}{Employed}                             & 10             & 53\%                 \\
\multicolumn{1}{r|}{Self-employed}                            & 2              & 11\%                 \\
\multicolumn{1}{r|}{Not working }        & 5              & 26\%                 \\
\multicolumn{1}{r|}{A student}                                & 2              & 11\%                 \\
\multicolumn{1}{r|}{A homemaker }                                 & 0              & 0\%                  \\ 
\multicolumn{1}{r|}{Prefer not to answer} & 0& 0\% \\
\midrule
\multicolumn{3}{l}{\textbf{Income} }                                             \\ 
\midrule
\multicolumn{1}{r|}{Less than \$30,000}                       & 3              & 16\%                 \\
\multicolumn{1}{r|}{\$30,000 to under \$65,000}              & 7              & 37\%                 \\
\multicolumn{1}{r|}{\$65,000 or more }                        & 9              & 47\%  \\
\multicolumn{1}{r|}{Prefer not to answer} & 0& 0\% \\ \bottomrule
\end{tabular}
\end{table}

\clearpage

\section{Quantitative Study}
\label{App:DesignImagesinSurvey}
\begin{figure*}[h!]
    \centering
\begin{subfigure}{0.45\textwidth}
  \includegraphics[width=\linewidth]{figures/confBlong.png} 
  \caption{CLB:This warning uses the signal word "confirmed" and a green border.}
  \label{fig:clb}
\end{subfigure}\hfil
\begin{subfigure}{0.45\textwidth}
  \includegraphics[width=\linewidth]{figures/unconf1.png}
  \caption{UcSB:This warning uses the signal word "confirmed" and a red border.}
  \label{fig:UcSB}
\end{subfigure}
\medskip
    \centering
\begin{subfigure}{0.45\textwidth}
  \includegraphics[width=\linewidth]{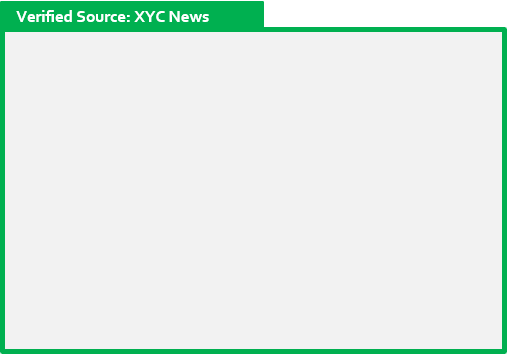} 
  \caption{vsc:This warning uses the signal word "verified" and a green border.}
  \label{fig:vsb}
\end{subfigure}\hfil 
\begin{subfigure}{0.45\textwidth}
  \includegraphics[width=\linewidth]{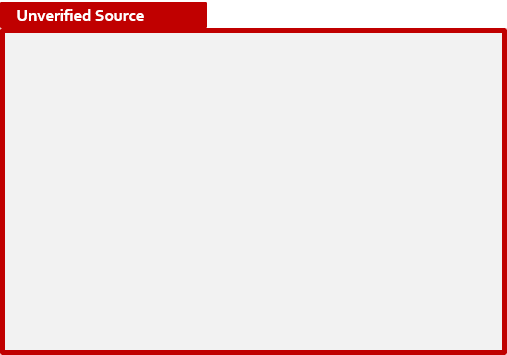}
  \caption{UvSB:This warning uses the signal word "verified" and a red border.}
  \label{fig:unsb}
\end{subfigure}
\caption{This figure shows the four designs we tested that have a border around the video instead of an icon.}
\label{fig:borderdesigns}
\end{figure*}

\begin{figure*}[h!]
    \centering
\begin{subfigure}{0.45\textwidth}
  \includegraphics[width=\linewidth]{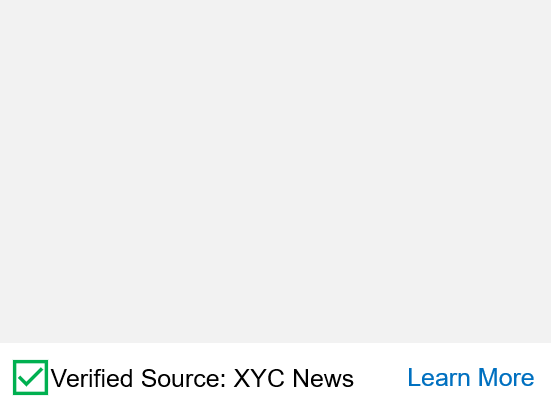} 
  \caption{VSC:This warning uses the signal word "verified" and a green check mark.}
  \label{fig:vsc}
\end{subfigure}\hfil 
\begin{subfigure}{0.45\textwidth}
  \includegraphics[width=\linewidth]{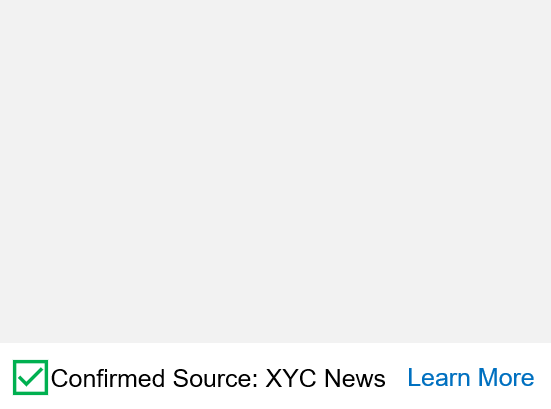} 
  \caption{CSC:This warning uses the signal word "confirmed" and a green check mark.}
  \label{fig:csc}
\end{subfigure}
\medskip
    \centering
\begin{subfigure}{0.45\textwidth}
\includegraphics[width=\linewidth]{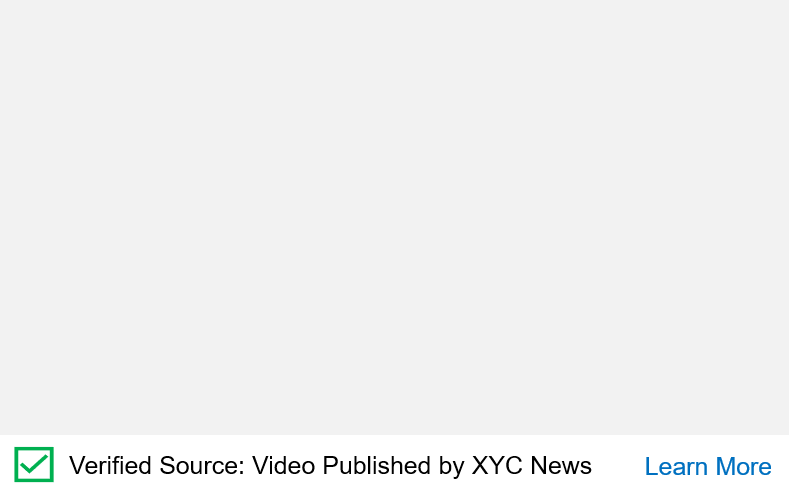} 
  \caption{VLC:This warning uses the signal word "verified", longer phrase and a green check mark.}
    \label{fig:vlc}
\end{subfigure}\hfil 
\begin{subfigure}{0.45\textwidth}
  \includegraphics[width=\linewidth]{figures/clc.png} 
  \caption{CLC:This warning uses the signal word "confirmed", longer phrase and a green check mark.}
  \label{fig:6}
\end{subfigure}
\caption{This figure shows the four designs that included the green check mark in their design to alert users that the source was validated.}
\label{fig:greencheckmarks}
\end{figure*}

\begin{figure*}[h!]
    \centering
\begin{subfigure}{0.45\textwidth}
  \includegraphics[width=\linewidth]{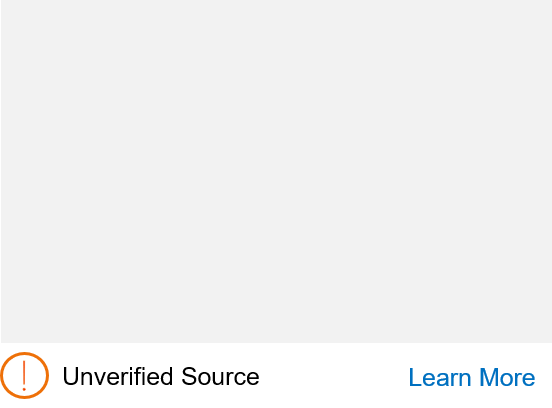} 
  \caption{UvSo:This warning uses the signal word "unconfirmed", short phrase and an exclamation point}
  \label{fig:uvso}
\end{subfigure}\hfil 
\begin{subfigure}{0.45\textwidth}
  \includegraphics[width=\linewidth]{figures/uso.png} 
  \caption{UcSo:This warning uses the signal word "unconfirmed", short phrase and an exclamation point}
  \label{fig:ucso}
\end{subfigure}

\caption{This figure displays the two designs that use the exclamation point icon in the warning to alert users that source was not validated.}
\label{fig:exclamdesigns}
\end{figure*}

\begin{figure}[h!]
     \centering
     \includegraphics[scale =.5]{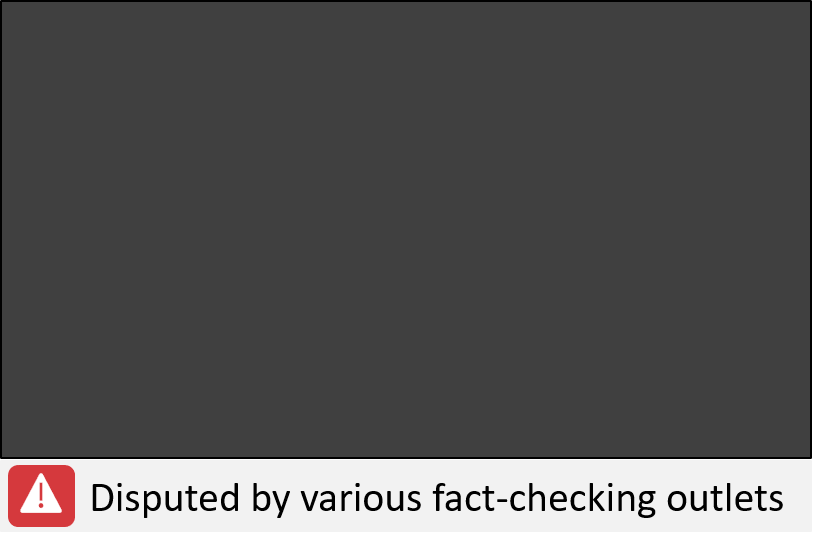}
     \caption{Facebook Misinformation Warning from 2017}
     \label{fig:fbwarning}
 \end{figure}

We next provide details of the quantitative study. Figure  \ref{fig:fbwarning} \ref{fig:borderdesigns}, \ref{fig:greencheckmarks}, and \ref{fig:exclamdesigns} display screenshots of the designs that were created based on participant and expert feedback. Table \ref{table:demoquant} indicates the demographics of the participants from the quantitative study where participants interpreted the designs.\\

\medskip

\begin{table}[hbt!]
\centering
\caption{Quantitative Study Demographics}
\label{table:demoquant}
\begin{tabular}{@{}r|cc@{}}
\toprule
\multicolumn{1}{r|}{\textbf{Demographics}} & \multicolumn{2}{c|}{\textbf{Participants}} \\ \midrule
\multicolumn{1}{r|}{(N=1,456)} & n & \% \\ \midrule
\multicolumn{3}{l}{\textbf{Gender}} \\ \midrule
Female & 588 & 40\% \\
Male & 863 & 59\% \\
Non-binary & 1 & \textless{}1\% \\
Prefer not to answer & 4 & \textless{}1\% \\ \midrule
\multicolumn{3}{l}{\textbf{Age}} \\ \midrule
18-30 & 540 & 37\% \\
31-40 & 481 & 33\% \\
41-50 & 243 & 17\% \\
Over 50 & 187 & 13\% \\
Prefer not to answer & 5 & \textless{}1\% \\ \midrule
\multicolumn{3}{l}{\textbf{Race}} \\ \midrule
White & 1049 & 72\% \\
Hispanic or Latino & 58 & 4\% \\
Black or African American & 243 & 17\% \\
Native American or American  Indian & 48 & 3\% \\
Asian & 100 & 7\% \\
Native Hawaiian or Pacific  Islander & 3 & \textless{}1\% \\
Other & 5 & \textless{}1\% \\
Prefer not to answer & 6 & \textless{}1\% \\ \midrule
\multicolumn{3}{l}{\textbf{Education}} \\ \midrule
No college & 90 & 6\% \\
Some college but no degree & 118 & 8\% \\
Associate's degree & 86 & 6\% \\
Bachelor's degree & 908 & 62\% \\
Advanced degree & 250 & 17\% \\
Prefer not to answer & 4 & \textless{}1\% \\ \midrule
\multicolumn{3}{l}{\textbf{Employment}} \\ \midrule
Employed & 1086 & 75\% \\
Self-employed & 234 & 16\% \\
Not working & 75 & 5\% \\
A homemaker & 32 & 2\% \\
A student & 22 & 2\% \\
Prefer not to answer & 7 & \textless{}1\% \\ \midrule
\multicolumn{3}{l}{\textbf{Income}} \\ \midrule
Less than \$30,000 & 324 & 22\% \\
\$30,000 to under \$40,000 & 678 & 47\% \\
\$65,000 or More & 445 & 31\% \\
Prefer not to answer & 9 & 1\% \\ \bottomrule
\end{tabular}%
\end{table}

\clearpage
\end{document}